\documentclass[a4paper,11pt]{article}
\pdfoutput=1 

\usepackage{jcappub} 

\usepackage[T1]{fontenc} 

\usepackage{amsmath,amssymb,amsfonts}
\usepackage{graphicx}
\usepackage{enumerate} 
\usepackage{colordvi} 
\usepackage{bm}
\usepackage{epsfig}
\usepackage{float}
\usepackage{verbatim}
\usepackage{subfig}
\usepackage{xcolor}
\usepackage{multirow,multicol,boldline}

\usepackage[colorinlistoftodos]{todonotes}
\usepackage{hyperref}

\newcommand{\bfk}{\mbox{\boldmath$k$}}
\newcommand{\bfp}{\mbox{\boldmath$p$}}

\newcommand{\cb}{\color{black}}

\newcommand{\cc}[1]{\textcolor{black}{#1}}

\usepackage[normalem]{ulem}

\title{\cc{Modelling the  matter bispectrum at small scales in modified gravity}} 

\author[a]{Benjamin Bose,}
\author[a]{Joyce Byun,}
\author[a,c]{Fabien Lacasa,}
\author[a,b]{Azadeh Moradinezhad Dizgah,}
\author[a]{Lucas Lombriser}

\affiliation[a]{D\'epartement de Physique Th\'eorique, Universit\'e de Gen\`eve, 24 quai Ernest Ansermet, 1211 Gen\`eve 4, Switzerland} 
\affiliation[b]{Department of Physics, Harvard University, 17 Oxford Street, Cambridge, MA 02138, USA}
\affiliation[b]{Universit\'e Paris-Saclay, CNRS, Institut d'astrophysique spatiale, 91405, Orsay, France}

\emailAdd{benjamin.bose@unige.ch}

\abstract{
Future large-scale structure surveys will measure three-point statistics with high statistical significance. This will offer significant improvements on our understanding of gravity, provided we can model these statistics accurately. \cc{We assess the performance of several schemes for theoretical modelling of the matter bispectrum, including halo-model based approaches and fitting formulae. We compare the model predictions against N-body simulations, considering scales up to $k_{\rm max} = 4 h/{\rm Mpc}$, well into non-linear regime of structure formation. Focusing on the equilateral configuration, we conduct this analysis for three theories of gravity: general relativity, $f(R)$ gravity, and the DGP braneworld model. Additionally, we compute the lensing convergence bispectrum for these models. We find that all current modelling prescriptions in modified gravity, in particular for theories with scale-dependent linear growth, fail to attain the accuracy required by the precision of the Stage IV surveys such as \emph{Euclid}. Among these models, we find that a halo-model corrected fitting formula achieves the best overall performance.}
}

\begin{document}
\maketitle
\flushbottom

\section{Introduction}
The standard model of cosmology, $\Lambda$CDM, has been hugely successful in reproducing many cosmological observations such as the cosmic microwave background (CMB)~\cite{Aghanim:2018eyx} and the large-scale structure (LSS) of the Universe~\cite{Anderson:2012sa,Song:2015oza,Beutler:2016arn}.
The model relies on two fundamental theoretical assumptions: that general relativity holds on all physical scales and that the Universe is statistically spatially homogeneous and isotropic. While $\Lambda$CDM fits observations extremely well overall, it also invokes two exotic dark components to be postulated, whose nature remains unexplained: cold dark matter (CDM) and dark energy in the form of a cosmological constant ($\Lambda$). Together, these account for $95\%$ of the matter-energy content of the Universe today. 
Despite the observational success of the concordance model, several mild tensions in cosmological parameters between late time measurements and the CMB have been uncovered with the increased precision of recent data.
In particular, there is a tension in the value of the Hubble parameter today, $H_0$,~\cite{Efstathiou:2013via,Zhang:2017aqn,Riess:2009pu} (cf.~\cite{Lombriser:2019ahl}) and in the amplitude of density fluctuations, $\sigma_8$,~\cite{Heymans:2013fya,Hildebrandt:2016iqg,Abbott:2017wau}
from direct measurement and that inferred from extrapolating the best-fit CMB data~\cite{Aghanim:2018eyx} (also see~\cite{Lin:2017bhs} for a review).
\\
\\
Motivated by these theoretical and observational issues, probing the nature of dark matter and dark energy, as well as testing alternatives to the standard model, is one of the main focuses of modern cosmology. A plethora of exotic dark energy and modified gravity models have been proposed over the past couple of decades for this purpose (for reviews, see~\cite{Copeland:2006wr,Clifton:2011jh,Joyce:2016vqv,Koyama:2018som}). 
However, any viable alternative to the standard model must pass all Solar System tests, match all cosmological data at least as well as $\Lambda$CDM and, moreover, not modify the speed of gravitational wave propagation~\cite{Lombriser:2015sxa,Monitor:2017mdv,Lombriser:2016yzn,Creminelli:2017sry,Ezquiaga:2017ekz,Baker:2017hug,Sakstein:2017xjx,Battye:2018ssx,deRham:2018red,Creminelli:2018xsv,Ishak:2018his}. This places very tight constraints on modifications to $\Lambda$CDM in the regimes of these experiments. 
\\
\\
One regime that still remains largely open to signals of modified gravity or exotic dark energy is within the LSS of the Universe, in particular, the non-linear, small physical scales of LSS.
These are much larger than the Solar-System scales but small enough to contain enough statistical information to discriminate between modifications to $\Lambda$CDM and other standard physics. For instance, it is in this regime that alternatives to $\Lambda$CDM are expected to give clear signatures as they transition from a modification of gravity on large scales, for example to effectively act as dark energy or dark matter, to recovering GR at Solar-System scales, where the gravitational interactions have been thoroughly tested.
In the context of modified gravity, this can be realised with screening mechanisms. These mechanisms come in three general flavours: models that screen through (i) deep potential wells or large (ii) first and (iii) second derivatives of the potential.
The chameleon~\cite{Khoury:2003rn} and symmetron~\cite{Hinterbichler:2010es}
mechanisms are of the first kind and exhibit a screening effect that is mass and environment dependent.
The second type is realised in k-mouflage models~\cite{Babichev:2009ee}, where screening occurs in regions of large acceleration.
Finally, the third type of screening, the Vainshtein mechanism~\cite{Vainshtein:1972sx}, operates in regions of high density.
\\
\\
Current and future LSS surveys such as 
KiDS\footnote{\url{http://kids.strw.leidenuniv.nl/}}~\cite{Kuijken:2015vca}, DES\footnote{\url{https://www.darkenergysurvey.org/}} \cite{Albrecht:2006um}, 
DESI\footnote{\url{https://www.desi.lbl.gov/}}
\cite{Aghamousa:2016zmz}, 
Euclid\footnote{\url{www.euclid-ec.org}} \cite{Laureijs:2011gra,Amendola:2016saw}, 
and LSST\footnote{\url{https://www.lsst.org/}} \cite{Chang:2013xja}
will measure the smaller scales of LSS with unprecedented precision and require equally precise theoretical modelling to describe the data.
On this note, much work has gone into developing 2-point matter and galaxy statistics for modified gravity models~\cite{Schmidt:2008tn,Koyama:2009me,Lombriser:2011zw,Li:2011uw,Hojjati:2011ix,Lombriser:2013aj,Lombriser:2013eza,Brax:2013fna,Zhao:2013dza,Lombriser:2014dua,Barreira:2014zza,Mead:2014gia,Lombriser:2015axa,Casas:2015qpa,Bose:2016qun,Mead:2016zqy,Lombriser:2016zfz,Bose:2017jjx,Bose:2017dtl,Aviles:2017aor,Hellwing:2017pmj,Cataneo:2018cic,Bose:2018orj,Winther:2019mus}, but studies are relatively limited for the 3-point statistics~\cite{Bose:2018zpk,Namikawa:2018erh,Hirano:2018uar,GilMarin:2011xq,Yamauchi:2017ibz,Bellini:2015wfa,An:2017kqu,Dinda:2018eyt,Munshi:2016zzr,Brax:2012sy} with most of the analytic works being restricted to a leading-order calculation, only valid at very large scales.
In the non-linear regime these studies have been largely restricted to simulations or phenomenological models, with the exception of \cite{Brax:2012sy}.
It is clear that the 3-point statistics will provide an invaluable means of further constraining cosmology and deviations from GR~\cite{Child:2018klv,Byun:2017fkz,Song:2015gca,Bose:2018zpk} and can be used in tandem with the power spectrum. For this reason, an accurate, gravity-general,
non-linear theoretical prescription should be provided to large-scale structure survey pipelines.
\\
\\
Our objective in this work is to assess current non-linear theoretical models for the matter bispectrum in modified gravity theories.
We take a chameleon screened and Vainshtein screened model as representatives of general scalar-tensor modifications, namely the Hu-Sawicki~\cite{Hu:2007nk} $f(R)$ model and the DGP~\cite{Dvali:2000hr} braneworld model (in its scalar-tensor limit).
These are well studied toy models for modified gravity~\cite{Clifton:2011jh,Joyce:2016vqv,Koyama:2018som}, which have been studied in detail with 
$N$-body simulations~\cite{2018slss.book.....L} and abundantly tested against observations~\cite{Koyama:2018som,Lombriser:2009xg,Lombriser:2014dua,Baker:2019gxo}.
They cover both potential and derivative screening, encompassing the effects of cluster mass, environment, and density on the behavior of the screening mechanism.
Additionally, $f(R)$ gravity exhibits a scale dependent growth of structure on linear scales, which further complicates the modelling of the bispectrum.
In the absence of a general fitting formula for the bispectrum that would encompass the variety of modified gravity and dark sector models, we turn to the halo model~\cite{Cooray:2002dia}, which currently seems like the most general non-linear approach for structure formation.
We will find here that compared to the level of accuracy achieved for the power spectrum, the halo model predictions for the modified gravity matter bispectrum are severely lacking. Improvements will be left to future work, but avenues for such improvement are open in light of recent developments on modelling modified gravity corrections to the GR power spectrum~\cite{Cataneo:2018cic}. Since we are interested in the small scale effects of modified gravity here, we also inspect the lensing convergence bispectrum, which is an integrated effect over the matter bispectrum and includes information from all scales.
\\
\\
Before moving on, we note that in~\cite{Brax:2012sy} the authors compare various non-linear models for the matter bispectrum, including the halo model, in a very
theoretical context. {\cb The authors do not compare these prescriptions for the bispectrum to simulations, which is the step this work makes. They also neglect screening largely, both in the 1-loop and spherical collapse calculations. Here, we include screening effects.} Furthermore, we compare variants of the halo model as well as halo model  corrections to $\Lambda$CDM fitting formulae, provide a full 1-loop bispectrum calculation in the relevant theoretical predictions, and investigate the lensing convergence bispectrum. 
\\
\\
The paper is organised as follows. In section~\ref{sec:theory} we briefly review the theoretical frameworks that will be employed for our predictions: perturbation theory, the halo model and a fitting formula. In section~\ref{sec:simcomp} we compare the predictions of these frameworks
against $N$-body simulations. In section~\ref{sec:lenscomp} we investigate the impact of non-linearities coming from modifications to gravity on the lensing bispectrum. Finally, in section~\ref{sec:summary} we summarise our results and conclude. 


\section{Theory: perturbations, halo model, fits and lensing convergence}\label{sec:theory}

We briefly review the theoretical aspects involved in computing the non-linear matter bispectrum for generalised theories of gravity.
In particular, we consider three different frameworks that can be used to predict the clustering statistics beyond the linear regime. We start in the quasi non-linear regime with standard Eulerian perturbation theory (SPT) in section~\ref{subsec:pt}, then extend to the non-linear regime with the halo model in section~\ref{subsec:hm}. Finally, in section~\ref{subsec:fit} we look at a recently proposed ansatz~\cite{Namikawa:2018erh} which takes information from $N$-body simulations to make its predictions. We note that the first two frameworks are based on fundamental theory and so are more flexible in terms of theoretical generality regarding gravity and the dark sector, whereas the last prescription is more restricted, only being valid for a subset of the general scalar-tensor theories of the Horndeski \cite{Horndeski:1974wa} and beyond-Horndeski~\cite{Gleyzes:2014qga,Gleyzes:2014dya,Langlois:2015cwa,Langlois:2015skt} class. 
\\
\\
Although all the approaches described in this section are completely gravity and dark energy general,  one of our main objectives is the comparison of the theoretical predictions with simulations. Thus, we must first choose specific models for which to compare. We take the DGP~\cite{Dvali:2000hr} and Hu-Sawicki~\cite{Hu:2007nk} $f(R)$ gravity models for which $N$-body simulations are readily available~\cite{2018slss.book.....L}.
These represent two distinct classes of modified gravity models: Vainshtein screened~\cite{Vainshtein:1972sx} and chameleon screened~\cite{Khoury:2003rn}.


\subsection{Perturbation theory}\label{subsec:pt}
Within SPT we can solve the momentum and energy conservation equations order by order to obtain higher order corrections to the  quantities of interest.  Working at larger scales, but far inside the Hubble horizon, we can treat the density field, $\delta$, perturbatively (see~\cite{Bernardeau:2001qr} for a review). Here we can safely apply the quasi-static approximation (see, however,~\cite{Lombriser:2015cla} for complications in beyond-Horndeski models).  In particular we make the assumption that the non-linear density and velocity perturbations are given as 
\begin{equation}
\delta_{NL}(\bfk;a) = \sum^{\infty}_{n=1}\delta_n(\bfk;a), \quad \quad \theta_{NL}(\bfk;a) = \sum^{\infty}_{n=1}\theta_n(\bfk;a),
\end{equation}
with 
\begin{align}
     \delta_n(\boldsymbol{k};a) \sim & \int d^3\boldsymbol{k}_1...d^3 \boldsymbol{k}_n \delta_D(\boldsymbol{k}-\boldsymbol{k}_{1...n}) F_n(\boldsymbol{k}_1,...,\boldsymbol{k}_n;a) \delta_0(\boldsymbol{k}_1)...\delta_0(\boldsymbol{k}_n),\label{densitypt} \\ 
          \theta_n(\boldsymbol{k};a) \sim & \int d^3\boldsymbol{k}_1...d^3 \boldsymbol{k}_n \delta_D(\boldsymbol{k}-\boldsymbol{k}_{1...n}) G_n(\boldsymbol{k}_1,...,\boldsymbol{k}_n;a) \delta_0(\boldsymbol{k}_1)...\delta_0(\boldsymbol{k}_n). 
\end{align}
$a$ is the scale factor, $\delta_n$ denotes the $n^{\rm th}$ order perturbation, $\delta_0$ is the primordial linear density perturbation, $\delta_D$ is the Dirac delta function and $k_{\rm 1...n}= k_1+...+k_n$. $F_n$ is the $n^{\rm th}$ order kernel function.\footnote{Note that $F_1$ is simply the growth factor, $D(a)$, under the Einstein-de Sitter (EdS) approximation, the time dependence of which is still determined by the modification to gravity (see \cite{Bose:2018orj} for example).} {\cb In Appendix~\ref{app:modgpt} we outline the process to calculate the $F_i$ in general theories of gravity and dark energy. }
\\
\\
{\cb We look to calculate the first order corrections to the linear matter statistics, the so called `1-loop' corrections. }In particular, for the 1-loop correction to the power spectrum we must solve up to the $3^{\rm rd}$ order density perturbation ($\delta_3$) and for the 1-loop bispectrum, we must solve up to the $4^{\rm th}$ order density perturbation ($\delta_4$). The total 1-loop-corrected matter power spectrum and bispectrum are given as  
\begin{align}
P^{{\rm loop}}(k;a) = & F_1^2(k;a)P_0(k) + [P^{22}(k;a) + P^{13}(k;a)], \label{1loopps} \\ 
B^{{\rm loop}}(k_1,k_2,\mu;a) = &  B^{112}(k_1,k_2,\mu;a) \nonumber \\ & + [B^{222}(k_1,k_2,\mu;a)  +  B^{321}(k_1,k_2,\mu;a) + B^{114}(k_1,k_2,\mu;a)] \label{1loopbs},
\end{align}
where $P_0(k)$ is the initial linear power spectrum and $\mu = (\hat{\bfk}_1\cdot \hat{\bfk}_2)$.\footnote{Not to be mistaken for the linear modification to gravity, $\mu(k;a)$, which will always include its arguments.} The tree level bispectrum ($B^{112}$) and 1-loop terms (shown in square brackets) are defined using the density field perturbations up to fourth order
\begin{align}
\langle \delta_{n_1}(\bfk_1) \delta_{n_2}(\bfk_2)\rangle &=
(2\pi)^3\delta_{\rm D}(\bfk_1+\bfk_2)\,P^{n_1 n_2}(k_1), \\ 
\langle \delta_{n_1}(\bfk_1) \delta_{n_2}(\bfk_2) \delta_{n_3}(\bfk_3) \rangle &=
(2\pi)^3\delta_{\rm D}(\bfk_1+\bfk_2 + \bfk_3 )\,B^{n_1n_2n_3}(\bfk_1,\bfk_2,\bfk_3), 
\end{align}
where we must add all permutations of the $n^{\rm th}$ density field perturbations, $\delta_n$, on the left-hand side. If we assume Gaussian initial conditions and under the usual assumptions of SPT, these averages can be decomposed into a product of linear power spectra convolved with the perturbative kernels, $F_i$. \cc{In Appendix~\ref{app:sptexp} we include the explicit integral expressions for the various terms in eq.~\eqref{1loopps} and eq.~\eqref{1loopbs}}, and direct the reader to \cite{Bose:2016qun,Bose:2018zpk} for a full description of this procedure for general models of gravity and exotic dark energy models, including the DGP and $f(R)$ models considered in this paper.


\subsection{Halo model}\label{subsec:hm}

 Next we summarise the key expressions of the halo model (see \cite{Cooray:2002dia} for a comprehensive review) in general theories of gravity \cite{Schmidt:2008tn,Schmidt:2009yj,Lombriser:2013eza,Barreira:2014zza,Lombriser:2014dua,Mead:2016zqy,Cataneo:2018cic}. This formalism assumes that matter is confined to halos, whose key properties determine the clustering statistics on all scales. As with perturbation theory, this framework is very general. {\cb In Appendix~\ref{app:sphercol} we describe the general collapse of  spherically symmetric over-densities into halos. By solving the collapse equations described in that appendix, one can obtain the halo model predictions for the matter power spectrum and bispectrum. }
 \\
 \\
 In order to solve the collapse equations and derive the matter statistics, one must also assume some forms for the basic physical properties of the halos. The final predictions  for LSS will be sensitive to these choices, and so here we consider a couple of well studied choices for these ingredients. 

\subsubsection{Halo model ingredients}\label{sec:haloingred}
To begin constructing any matter statistics there are a few quantities that we need to know or provide a prescription for if we want to understand correlations within the halos themselves. These are 
\begin{enumerate}
    \item 
    The number density or abundance, $n_{\rm vir}(M_{\rm vir})$, of halos of a given mass, $M_{\rm vir}$. This is directly related to the {\it mass function}.
    \item
    The {\it density profile} of these halos, $\rho_h(r,M_{\rm vir})$.  
       \item 
    A measure of how concentrated mass is within the halo. This is usually parametrised using the {\it concentration parameter}, $c_{\rm vir}(M_{\rm vir})$.
\end{enumerate}
Note that all the essential quantities $n_{\rm vir}$, $\rho_h$ and  $c_{\rm vir}$ are time dependent, and this is assumed implicitly to avoid clutter. In this work, we will assume some popular ans\"atze for modelling these ingredients that have been shown to work well for GR.
Importantly, with some reinterpretation of the input variables, they have also been shown to provide accurate descriptions for these quantities in DGP and $f(R)$ gravity by tests against $N$-body simulations~\cite{Schmidt:2008tn,Schmidt:2009sv,Lombriser:2012nn,Lombriser:2013wta,Lombriser:2013eza,Barreira:2014zza,Cataneo:2016iav,Cataneo:2018cic,Mitchell:2019qke}. Our choices will aim to determine how general these ingredients can be, and if there is  a need to finely tune  them to particular simulation measurements. 
\\
\\
The halo mass function is given by 
\begin{equation}
      n_{\rm vir} \equiv \frac{ {\rm d} n}{{\rm d}\ln{M_{\rm vir}}} = f(\nu)\frac{\bar{\rho}_{m,0}}{M_{\rm vir}}\frac{ {\rm d} \nu}{{\rm d}\ln{M_{\rm vir}}}, \label{eq:nvir}
\end{equation}
where we have defined the peak threshold $\nu \equiv \delta_c/\sigma$,\footnote{{\cb See Appendix~\ref{app:sphercol} on how to calculate $\delta_c$.}} $\sigma$ being the variance of the linear density field smoothed with a top hat of comoving radius $R_{\rm vir}$ 
\begin{equation}
    \sigma^2(R_{\rm vir},a) = \int \frac{d^3k}{(2\pi)^3} |W(kR_{\rm vir})|^2 P_L(k;a). \label{sigmaeq}
\end{equation}
$W(k)$ represents the Fourier transform of a top-hat filter and $P_L(k;a)$ is the linear power spectrum evolved to the scale factor $a$. We will consider two separate forms for the multiplicity function $f(\nu)$. The Sheth-Tormen ansatz \cite{Sheth:1999mn,Sheth:2001dp} and the more recent Tinker \emph{et al}.~\cite{Tinker:2010my} ansatz: 
\begin{align}
    f^{\rm ST}(\nu) & = \frac{1}{\nu} \left( A \sqrt{\frac{2}{\pi}q\nu^2} \left[ 1+(q\nu^2)^{-p}\right] \exp{\left[-q\nu^2/2 \right]} \right), \label{eq:nvirst} \\ 
    f^{\rm T}(\nu) & = \alpha \left[1+(\beta \nu^2)^{-2\phi}\right] \nu^{4\eta} \exp{(-\gamma \nu^4/2)}. \label{eq:nvirt}
\end{align}
The  parameters of the ans\"atze $f^{\rm ST}(\nu)$ and $f^{\rm T}(\nu)$ are calibrated to $\Lambda$CDM simulations. {\cb We list the numerical fits for these in Appendix~\ref{app:bias}, eq.~\eqref{bolshoiconst}.} Note that the Tinker mass function fits assume $\Delta_{\rm vir} = 200$ (see Appendix~\ref{app:sphercol}, eq.~\eqref{deltavir}) and so we impose this when using this expression. Furthermore, the constants $A=0.322$ and $\alpha=0.368$ are normalisation constants obtained by imposing that all mass in the Universe is contained in halos, i.e. $\int d\nu f(\nu) = 1$. 
\\ 
\\ 
For the density profile we take the Navarro-Frenk-White (NFW) profile \cite{Navarro:1996gj} 
\begin{equation}
    \rho_h(r) = \frac{\rho_s}{r/r_s (1+r/r_s)^2}, \label{eq:nfw}
\end{equation}
where $r_s= R_{\rm vir}/c_{\rm vir}$. $c_{\rm vir}$ is the concentration parameter (see eq.~\eqref{eq:cvir} and eq.~\eqref{eq:bol}) and 
\begin{equation}
    \rho_s = \frac{M_{\rm vir}}{4\pi r_s^3}\left[\ln{(1+c_{\rm vir})}-\frac{c_{\rm vir}}{1+c_{\rm vir}} \right]^{-1}.
\end{equation}
As for the mass function, we consider two forms for the concentration parameter: a simple power-law expression~\cite{Bullock:1999he} and a functional fit to the Bolshoi $\Lambda$CDM simulation~\cite{Klypin:2010qw}: 
\begin{align}
    c^{\rm PL}_{\rm vir}(M_{\rm vir}) =& c_0 a \left(\frac{M_{\rm vir}}{M_\star}\right)^{-\alpha_0},\label{eq:cvir} \\ 
       c^{\rm Bol}_{\rm vir}(M_{\rm vir}) =& \lim_{k\to 0}9.2 \kappa_c(a) F_1^{1.3}(k;a)\left( \frac{M_{\rm vir}}{10^{12}}\right)^{-0.09} \left[1+0.013\left(\frac{M_{\rm vir}}{10^{12}} F_1^{-14.44}(k;a)\right)^{0.25} \right],\label{eq:bol}
\end{align}
where $c_0=9$ and $\alpha_0 =0.13$, which are also obtained from fits to $\Lambda$CDM simulations.
The characteristic mass $M_\star$ is found by solving $\nu(M_\star)=1$.
Note that due to the chameleon effect, in $f(R)$ gravity the extrapolated $M_\star$ will vary for different halo masses~\cite{Lombriser:2012nn,Lombriser:2013eza}.
Furthermore, we adopt $\kappa_c(a) = 1.26$ for $a=1$ and 0.96 for $a\leq 0.5$ as in~\cite{Giannantonio:2011ya}. In $f(R)$ we find that introducing the scale dependence to the growth factor $F_1$ in eq.~\eqref{eq:bol} deteriorates the fit of $c^{\rm Bol}_{\rm vir}(M_{\rm vir})$ to the simulations and so we take the large scale limit as indicated. Comparing the Bolshoi concentration relation to the general power law will give an indication of the importance of accurate concentration relations in modified gravity theories. On this note, we remark that whereas in the power-law relation screening effects enter through $M_\star$ via $\nu$, the Bolshoi fit includes no such screening information.


\subsubsection{Halo model bispectrum and power spectrum} \label{sec:halobipo}
Using the halo model, we can construct the 3-point correlation statistic of matter, which has three contributions, the so-called 1-, 2- and 3-halo terms, which correspond respectively to having all three points within the same halo, two points in the same halo and the third in a different halo, and three points in three different halos. Statistics between triplets of halos can be modeled using the tree level prediction, $B_{112}$, or any quasi non-linear prescription, for example the 1-loop bispectrum given in eq.~\eqref{1loopbs}.
\\
\\
Thus, the total halo model power spectrum is given by (see for example \cite{Cooray:2002dia}) 
\begin{equation}
        P_{\rm HM}(k) = P^{\rm 2h}(k)+  P^{\rm 1h}(k), \label{eq:halomps}
\end{equation}
while the total halo model bispectrum is given by (see for example \cite{Cooray:2002dia,Lacasa:2013yya})
\begin{equation}
    B_{\rm HM}(k_1,k_2,\mu) = B^{\rm 3h}(k_1,k_2,\mu) + B^{\rm 2h}(k_1,k_2,\mu) + B^{\rm 1h}(k_1,k_2,\mu). \label{eq:halom}
\end{equation}
We have dropped the time dependence within the arguments in the expressions for simplicity. The individual terms are given in appendix~\ref{app:bias}.
\\
\\ 
Finally, in our analysis we will consider three different sets of halo model ingredients: 
\begin{itemize}
    \item 
    The Sheth-Tormen mass function eq.~\eqref{eq:nvirst} with the power-law virial concentration given in eq.~\eqref{eq:cvir}, denoted {\bf ST}.
    \item
    The Tinker mass function in eq.~\eqref{eq:nvirt} with the Bolshoi fit concentration in eq.~\eqref{eq:bol} as used in \cite{Lazanu:2015rta}, denoted as {\bf Laz}.
    \item
    The Tinker mass function with the power-law concentration, denoted as {\bf Tink}.
\end{itemize}
\cc{We summarise these models again at the beginning of section~\ref{sec:simcomp}. We will also denote with a `-T' and `-L' whether we use the tree level or 1-loop computation for the 2-halo term in the power spectrum and 3-halo term in the bispectrum (see eq.~\eqref{2hps} and eq.~\eqref{3hbs}). So for example ST-L, with reference to the bispectrum say, will indicate a halo model prediction using the Sheth-Torman mass function, power-law virial concentration and a 1-loop prediction in the 3-halo term.}

\subsection{\cc{Fitting formulae and corrections}}\label{subsec:fit}
In \cite{Scoccimarro:2000ee,GilMarin:2011ik}, the authors present a formula for the matter bispectrum which is fit to $\Lambda$CDM simulations 
\begin{equation}
B^{\rm fit}(k_1,k_2,\mu;a) = 2 F^{fit}_2(\bfk_1,\bfk_2;a) P_{NL}(k_1;a) P_{NL}(k_2;a)  + {\rm 2 \; permutations \; including \;} k_3,  \label{eq:bfit} 
\end{equation}
 where $P_{NL}$ is the non-linear matter power spectrum, which can be calculated from a prescription of choice. As in~\cite{GilMarin:2011ik}, we will employ the  halo-model-based  halofit formula~\cite{Smith:2002dz,Takahashi:2012em,Smith:2018zcj} ({\bf HF}) for $P_{NL}$. The kernel $F^{fit}$ includes the fits to simulations and is given as \cite{Scoccimarro:2000ee}
\begin{align}
F^{\rm fit}_2(\bfk_1,\bfk_2;a) = &  F_1(a)^2 \Big[\left(\kappa(a)- \frac{2}{7}\lambda(a)\right)\bar{a}(k_1,a)\bar{a}(k_2,a) \nonumber \\ & \qquad  \qquad + \frac{\kappa(a)}{2} \mu \frac{k_1^2 + k_2^2}{k_1 k_2}\bar{b}(k_1,a)\bar{b}(k_2,a)  + \lambda(a) \frac{2}{7} \mu^2\bar{c}(k_1,a)\bar{c}(k_2,a) \Big].
\label{fittingbh}
\end{align}
We detail the non-linear functions $\bar{a}$, $\bar{b}$ and $\bar{c}$ in appendix~\ref{app:gmform}. \cc{These functions are fit to $\Lambda$CDM simulations. Specifically, we use the fits found in \cite{GilMarin:2011ik}}. In \cite{Scoccimarro:2000ee,GilMarin:2011ik} $\lambda(a) = \kappa(a) = 1$, and non-unity values will be discussed next. We will refer to the case of $\kappa(a) = \lambda(a) = 1$ as {\bf GM} after the authors of \cite{GilMarin:2011ik}. 
\\
\\
Now, we justify the introduction of the time dependent functions $\lambda(a)$ and $\kappa(a)$ in eq.~\eqref{fittingbh}. These were introduced into the fitted kernel for the matter bispectrum (eq.~\eqref{fittingbh}) in~\cite{Namikawa:2018erh}, and are motivated by a subset of the  Horndeski~\cite{Horndeski:1974wa} and beyond-Horndeski~\cite{Gleyzes:2014qga,Gleyzes:2014dya,Langlois:2015cwa,Langlois:2015skt} class of theories. These classes of theories constitute the broadest class of theoretically viable modifications to GR based on a single additional scalar degree of freedom in four dimensions. 
\\
\\
To see where $\lambda(a)$ and $\kappa(a)$ come from, we need only write down the second-order kernel in (beyond-)Horndeski theories under the quasi-static approximation
\begin{equation}
F_2(\bfk_1,\bfk_2;a) = F_1(a)^2 \Big[ \frac{\kappa(a)}{2} \left[ \alpha(\bfk_1,\bfk_2) + \alpha(\bfk_2,\bfk_1) \right] - \frac{2 }{7} \lambda(a) (1-\mu^2) \Big],
\label{horndeskif2}
\end{equation}
where again $\mu = (\hat{\bfk}_1 \cdot \hat{\bfk}_2$), $F_1(a)$ is the linear growth factor, $\alpha({\bfk}_1,{\bfk}_2)$ is the standard mode coupling kernel from SPT (see eq.~\eqref{alphabeta}) and $\kappa(a)$ and $\lambda(a)$ are second-order time dependent functions that are dependent on the specific theory under consideration. For Horndeski models we have $\kappa(a)=1$ and for GR $\kappa(a) =\lambda(a) =1$. Eq.~\eqref{horndeskif2} was the key motivation for the form of $F^{fit}$ presented in eq.~\eqref{fittingbh}. We will call eq.~\eqref{eq:bfit} for the case when $\kappa(a)$ or $\lambda(a)$ are non-unity, {\bf NBT}, after the authors of \cite{Namikawa:2018erh} who proposed this extension to the GM model. 
\\
\\
{\cb We now make a few notes on the validity of the NBT formula }
\begin{enumerate}
    \item 
\cb{The subset of theories which this fitting formula is valid for is that which exhibits a scale-independent linear growth of structure.} In particular, the formula proposed in~\cite{Namikawa:2018erh} is valid only for the subset of theories for which the scalar field potential term in the equations of motion is only time dependent (and not a function of scale). In this way one can adopt the Einstein-de Sitter (EdS) approximation:  the modification to the second-order perturbation theory kernel comes in the form of time dependent scalings of the EdS  second-order kernel. {\cb Note that  $f(R)$ gravity includes a scale dependent potential term which means the scale dependence of its second order kernel cannot be described by the scale dependence of the EdS kernel.  Since the fitting formula relies on the EdS kernel, the fitting formula will not be valid for $f(R)$, or more generally, any such theory that modifies the EdS kernels significantly.  In DGP, this is not the case, and the corrections to the EdS kernels are very small (see \cite{Bose:2018orj} for example).} We will however consider a fully generalised, halo-model-corrected  version of this formula  at  the end  of this  section (see eq.~\eqref{eq:correctedgm}).
\item
Recall that the functions $\bar{a}$, $\bar{b}$ and $\bar{c}$ in eq.~\eqref{fittingbh} \cc{ are fit to $\Lambda$CDM simulations. The impact of these fits, as well as the  ans\"atze for their functional forms, will introduce inaccuracies for beyond $\Lambda$CDM scenarios. This may not be of concern if we consider corrections to eq.~\eqref{eq:bfit}, given by the halo model for example. This is considered later in this section (see eq.~\eqref{eq:correctedgm}).} The NBT formula does however encode some screening information in the modification of the $F_2$ kernel given in eq.~\eqref{horndeskif2}. 
\item
For $P_{NL}$ in eq.~\eqref{eq:bfit}, we use the HF prescription but replace the linear growth factors of $\Lambda$CDM with the modified counterparts. By using the HF formula in eq.~\eqref{eq:bfit} we also incur inaccuracies in general theory predictions as HF is fit to $\Lambda$CDM simulations. These can also be corrected to an extent by applying some correction factor based on the halo model. We find that these inaccuracies are mild (see magenta and purple curves in figure~\ref{ps_dgp} for example) for scale independent theories such as DGP. 
\end{enumerate}
\cb{Based on these notes, we will only consider NBT for DGP gravity. We give explicit forms for $\lambda$ and $\kappa$ for DGP gravity in appendix~\ref{app:gmform}.}
\\
\\
Finally, for the power spectrum comparisons in this paper, we will also compare  HF. For modifications to GR one can apply a halo-model reaction~\cite{Cataneo:2018cic} to this formula to account for non-linear modifications to gravity. This method involves the 1- and 2-halo terms, combined with 1-loop perturbation theory, to compute a correction factor, $\mathcal{R}$, encoding non-linear modifications to gravity. \cc{The idea is that the halo model, while failing to model the  spectra, does well at modelling the modification to $\Lambda$CDM when we change the theory of gravity or dark energy. Based on this, one can apply a multiplicative factor to an accurate $\Lambda$CDM spectrum that captures this modification, that is essentially a ratio of halo model terms. $\mathcal{R}$ is slightly more sophisticated than a pure ratio of halo model terms. It includes extra ingredients that help model the transition regime between 1-halo and 2-halo terms. Further, it captures additional information on the modification in the quasi non-linear regime through its inclusion of 1-loop perturbation theory based ingredients.}
\\
\\
The halo-model reaction approach also  involves a  so  called non-linear `pseudo' power  spectrum that includes modified gravity effects only in the linear power spectrum. We refer the interested reader to appendix~\ref{app:reaction} for details, \cc{and just present the formula here}
\begin{equation}
    P^{\rm HF-\mathcal{R}} = \mathcal{R}(k) P^{\rm HF,pseudo}(k), \label{eq:react}
\end{equation}
\cc{where $P^{\rm HF,pseudo}$ is the halofit power spectrum computed using a linear power spectrum in the modified theory.}
This reaction corrected halofit formula will be denoted as {\bf HF-$\mathcal{R}$}.
\\
\\
Somewhat similarly, for the bispectrum comparisons, we will show eq.~\eqref{eq:bfit} as well as a halo-model corrected version of the GM fitting formula
\begin{equation}
    B^{\rm GM ,corrected} = B^{\rm fit, GM}_{\rm GR}\times \frac{B_{\rm HM,MG}}{B_{\rm HM,GR}}, \label{eq:correctedgm}
\end{equation}
where the subscript MG indicates the computation is done in a modified gravity theory (DGP or $f(R)$ in this work) and \cc{the subscript HM refers to a computation made using the halo model. Specifically, we use the Tink-T model described in sec~\ref{sec:halobipo}. The reason for choosing Tink-T above any other halo model variant is because it was found to outperform the others during our comparisons (seen in the next section). We will denote eq.~\eqref{eq:correctedgm} as {\bf GM-C}.} Since the bispectrum relies on three wave vectors, an ansatz for a sophisticated correction, such as the halo model reaction $\mathcal{R}$ (see eq.~\eqref{eq:reaction}), is non-trivial, and is left for a future work.


\subsection{Convergence bispectrum }\label{sec:lensing}
Outside of $N$-body simulations, one does not directly measure the matter bispectrum. Yet the matter bispectrum is the core ingredient for several observable quantities such as the galaxy clustering bispectrum (see~\cite{Gil-Marin:2014pva} for example) or the lensing convergence bispectrum.  The signatures of modified gravity are largest at scales larger than the screened regime but small enough so that fifth force effects are strong~\cite{Winther:2015wla}. Which scales this corresponds to exactly is highly model dependent, and so in this paper we focus our attention on weak lensing statistics, which in principle probe all scales. We note that galaxy clustering will also be an invaluable and complementary probe into gravity as it encodes valuable velocity information (for example see~\cite{Hellwing:2014nma,Li:2012by,Linder:2007nu,Guzzo:2008ac}), but this requires redshift space modelling as well as galaxy bias modelling. Because of this, clustering models are usually restricted to the quasi non-linear regime~\cite{Hashimoto:2017klo,Kaiser:1987qv,Desjacques:2018pfv}.
\\
\\
Gravitational lensing of background galaxies by the intervening matter distribution between the source and the observer, induces a magnification or convergence of the images of the source galaxies. The lensing convergence bispectrum gives a measure of the correlation in triplets of points in these convergence (magnification) maps. Therefore, the convergence bispectrum probes the bispectrum of intervening matter, projected along the line of sight.
\\
\\
Here we provide the theoretical expression for the weak lensing convergence bispectrum which requires a proper modelling of the matter bispectrum (see \cite{Cooray:2000uu} for a review). This is given in the flat-sky limit as 
\begin{equation}
    B_{\ell_1 \ell_2 \ell_3} = \int_0^{\chi_\star} {\rm d}\chi \left[\frac{3\Omega_{m,0} H_0^2}{2 a(\chi)} \right]^3 \chi^2 W^3(\chi,\chi_\star) B_{\delta}(\ell_1/\chi, \ell_2/\chi, \mu ; a(\chi)) \label{eq:lensbs}, 
\end{equation}
where $B_{\delta}$ denotes the matter bispectrum, $\chi$ is the comoving distance, $\chi_\star$ denotes the distance to the source and the window function $W$ is given by
\begin{equation}
    W(\chi,\chi_\star) = \frac{\chi_\star - \chi}{\chi \chi_\star}.
\end{equation}
Note that the form of the lensing spectrum is unaltered in the theories of gravity we consider (see \cite{Joyce:2016vqv} for a review on this topic). We will consider eq.~\eqref{eq:lensbs} in section~\ref{sec:lenscomp} once we determine the most accurate matter bispectrum models among the ones discussed here, through comparison against measurements from $N$-body simulations. This will be done next. 


\section{Matter spectra: comparisons to simulations}\label{sec:simcomp}
In this section we compare the theoretical predictions outlined in section~\ref{sec:theory} to measurements from $N$-body simulations. Before proceeding, we will give a summary of all the various models considered as well as all acronyms and abbreviations used. 
\\
\\
\cc{In table~\ref{model_table_1} we list the halo models we consider. These are outlined in section~\ref{subsec:hm} and each provides a prediction for both the power spectrum and bispectrum. Specifically, we consider three different versions of the halo model, each being constructed with a different combination of mass function, $n_{\rm vir}$, and virial concentration, $c_{\rm vir}$. Further, for each of these we consider two different forms for the perturbative components, $P^{\rm pt}$(eq.~\eqref{2hps}) and $B^{\rm pt}$ (eq.~\eqref{3hbs}). We will denote the tree level computation with a `-T' and the 1-loop computation with a `-L'. 
\\
\\
In table~\ref{model_table_2} we list various fitting formulae we consider for the power spectrum and bispectrum respectively, as well as the corrected fitting formulae for non-standard physics. In this table we include a brief description of the formula and of the correction factor.  
\\
\\
 We will not include the tree and 1-loop level modelling in the comparisons to avoid cluttering the plots. These typically fail at large scales and we are primarily interested in the small scale modelling. In appendix~\ref{app:pertvsnl} we show the comparison of SPT-based models against N-body as a reference. In this section, unless otherwise stated $P_{\rm ref}$ and $B_{\rm Ref}$ will refer to the $N$-body simulation measurements of the power spectrum and bispectrum respectively.
\\
\\
Finally, we remind the reader that all models listed in table.~\ref{model_table_1} as well as the fitting formula based HF-$\mathcal{R}$ and GM-C, are theoretically general and so are in principle applicable to beyond $\Lambda$CDM physics. NBT would be applicable to scale-independent theories within the Horndeski and beyond Horndeski class of scalar-tensor theories. 
}

\begin{table}[h]
\centering
\caption{Halo models for the power spectrum and bispectrum, indicating what forms are used for the mass function and virial concentration. The `-T' and '-L' in the halo-model models indicates the model uses the tree or 1-loop SPT prediction for the 2- and 3-halo term for the power spectrum and bispectrum respectively. The general form for the halo model power spectrum and bispectrum are given in eq.~\eqref{eq:halomps} and eq.~\eqref{eq:halom}. \cc{All of these models can encompass general modifications to $\Lambda$CDM}.}  
\begin{tabular}{| c || c | c | c | } \hline 
 \multicolumn{4}{| c |}{{\bf Halo Model} (eq.~\eqref{eq:halomps} and eq.~\eqref{eq:halom}) } \\ \hline 
 $c_{\rm vir}$   &  \multicolumn{2}{c|}{Power law} & Bolshoi  \\ \hline 
  $n_{\rm vir}$ &  Sheth-Tormen &   Tinker & Tinker  \\ \hline \hline 
 {\bf ST-T,ST-L} &  \checkmark & -   & -    \\ \hline
  {\bf Tink-T,Tink-L}  &  - & \checkmark  &  -  \\ \hline
  {\bf Laz-T,Laz-L } &  - &  -  &  \checkmark  \\ \hline
  \end{tabular}  
\label{model_table_1}
\end{table}

\begin{table}[h]
\centering
\caption{Fitting formulae used and a brief description.} 
\begin{tabular}{| c |} \hline 
 \multicolumn{1}{ |c|}{{\bf Fitting formulae for power spectrum}} \\ \hline \hline
 \multicolumn{1}{ |c|}{{\bf Halofit (HF) \cite{Takahashi:2012em} }} \\ \hline 
 \multirow{2}{\textwidth}{Functional form for $P(k)$ based on the halo model but with free parameters calibrated to $\Lambda$CDM simulations.}  \\ \\ \hline
  \multicolumn{1}{ |c|}{{\bf Halofit corrected formula (HF-$\mathcal{R}$) \cite{Cataneo:2018cic} }} \\ \hline 
   \multirow{2}{\textwidth}{HF formula multiplied by the reaction $\mathcal{R}$, which corrects for non-standard  physics. The reaction is calculable using SPT and the halo model. This is given by eq.~\eqref{eq:react}. } \\ \\ \hline 
       \multicolumn{1}{l}{} \\
    \hline
 \multicolumn{1}{ |c|}{{\bf Fitting formulae for bispectrum}} \\ \hline \hline
 \multicolumn{1}{ |c|}{{\bf Gil-Marin et al formula (GM) \cite{GilMarin:2011ik} }} \\ \hline 
 \multirow{2}{\textwidth}{Functional form for $B(k)$ based on the tree level bispectrum eq.~\eqref{tree2} formula but with additional scale dependencies calibrated to $\Lambda$CDM simulations. This is given by eq.~\eqref{eq:bfit} with $\kappa(a) = \lambda(a) =1$ in eq.~\eqref{fittingbh}.}  \\ \\ \\ \hline
 \multicolumn{1}{ |c|}{{\bf Namikawa et al formula (NBT) \cite{Namikawa:2018erh}}} \\ \hline 
 \multirow{3}{\textwidth}{Same as GM formula but with corrections to eq.~\eqref{fittingbh} coming from the Horndeski and beyond Horndeski class of theories in the form of general $\kappa(a)$ and $\lambda(a)$. This is given by eq.~\eqref{eq:bfit}. This does not encompass corrections for scale-dependent theories such as $f(R)$. }  \\ \\ \\ \hline
 \multicolumn{1}{ |c|}{{\bf GM-corrected formula (GM-C)}} \\ \hline 
 \multirow{2}{\textwidth}{Same as GM formula but multiplied by a ratio of halo model calculations to correct for non-standard physics. This is given by eq.~\eqref{eq:correctedgm}. }  \\ \\ \hline 
  \end{tabular}  
\label{model_table_2}
\end{table}
\newpage 

\subsection{Simulations and errors}
The simulations used are detailed in~\cite{Cataneo:2018cic} and we provide a brief summary here. They were run using {\tt ECOSMOG}~\cite{Li:2011vk,Li:2013nua} and are dark matter only, with $1024^3$ particles in a box of side length $512{\rm Mpc}/h$.
The initial conditions are generated using 2LPTic~\cite{Crocce:2006ve} at $z_{\rm ini} = 49$, and the phases for the initial density fields are the same for all simulations to reduce the effect of cosmic variance. 
The linear power spectrum, generated using CAMB~\cite{Lewis:1999bs}, has the following flat cosmology: $\Omega_{\rm m} =0.3072$, $\Omega_{\Lambda} = 0.6928$, $h=0.68$, $\Omega_{\rm b} = 0.0481$, $\sigma_8(z=0) = 0.8205$, and $n_s=0.9645$. 
\\
\\
The modified gravity simulations come with an additional parameter. For $f(R)$ gravity we have $|f_{\rm R0}|=10^{-5}$ (we denote this as $F5$ in plot legends and labels) and for the DGP simulation $H_0 r_c = 0.5$. \cc{These  parameter values exhibit somewhat larger modifications compared to the current constraints on these models (for example ~\cite{Barreira:2016ovx,Burrage:2017qrf}), and are large enough to provide a good test for the  theoretical modelling of modified gravity, which is a core objective of this work.} \cc{ For example, for DGP, $H_0 r_c > 2$ at $2\sigma$ from redshift space distortions \cite{Barreira:2016ovx}. On the other hand, for $f(R)$, the value we choose is comparable to current constraints from weak lensing \cite{Liu:2016xes}, where they constrain at the $10^{-5}$ level. Assuming a lower modification would bring the predictions closer to the GR predictions, which generally perform better as all models have some level of calibration to GR-based simulations}.
\\
\\
Further, as we are primarily interested in the late-time Universe, and in particular in the galaxy lensing, we consider the redshifts $z=0.2,0.5,1$ and $1.5$ for the $\Lambda$CDM and $f(R)$ cases. These redshifts were chosen based on the target source galaxies for future lensing surveys and target redshift range for galaxy clustering measurements, for example with the upcoming Euclid satellite~\cite{Laureijs:2011gra}. Note that for DGP, we only have snapshot data available at $z=0$ and 1, and so these comparisons are restricted.
\\
\\
To measure the power spectrum and bispectrum from these simulations we use FFT-based estimators with $N_{grid} = 1024$, combined with the interlacing algorithm from~\cite{Sefusatti:2015aex}. 
For the bispectrum, we measure only the equilateral triangle bins with bin centres from $k_{min}=6k_f$ to $k_{max}=336k_f$ with bin width $\Delta k = 6 k_f$, where $k_f \equiv 2\pi/L$.
\\
\\
Since we only have a single realisation for each model (GR, $f(R)$ and DGP), we choose to model the errors using the Gaussian variance \cite{Zhao:2013dza,Feldman:1993ky,Seo:2007ns} 
\begin{align}
    \sigma_p(k) =& \frac{2\pi P(k)}{k\sqrt{\Delta k V}}, \label{pserrs} \\ 
    \sigma_b(k_1,k_2,k_3) =& \sqrt{\frac{V s_b P(k_1) P(k_2) P(k_3)}{N_\Delta} }, \label{bserrs}
\end{align}
where $V = 512^3 {\rm Mpc}^3/h^3$, $P(k)$ is the non-linear matter power spectrum measured from the simulations, $N_\Delta$ is the number of fundamental triangles in the bin and $\Delta k$ is the bin width. We have $s_b=6$ since we only consider equilateral triangle bins, but it can otherwise be 2 or 1 for isosceles and scalene triangle bins, respectively. \cc{Note that we do not have a shot-noise term since we are only considering matter clustering}. In all plots we show the $2\sigma$ error bands.

\subsection{Power spectrum results}
\begin{figure}[htbp!]
\centering
  \includegraphics[width=\textwidth,height=8cm]{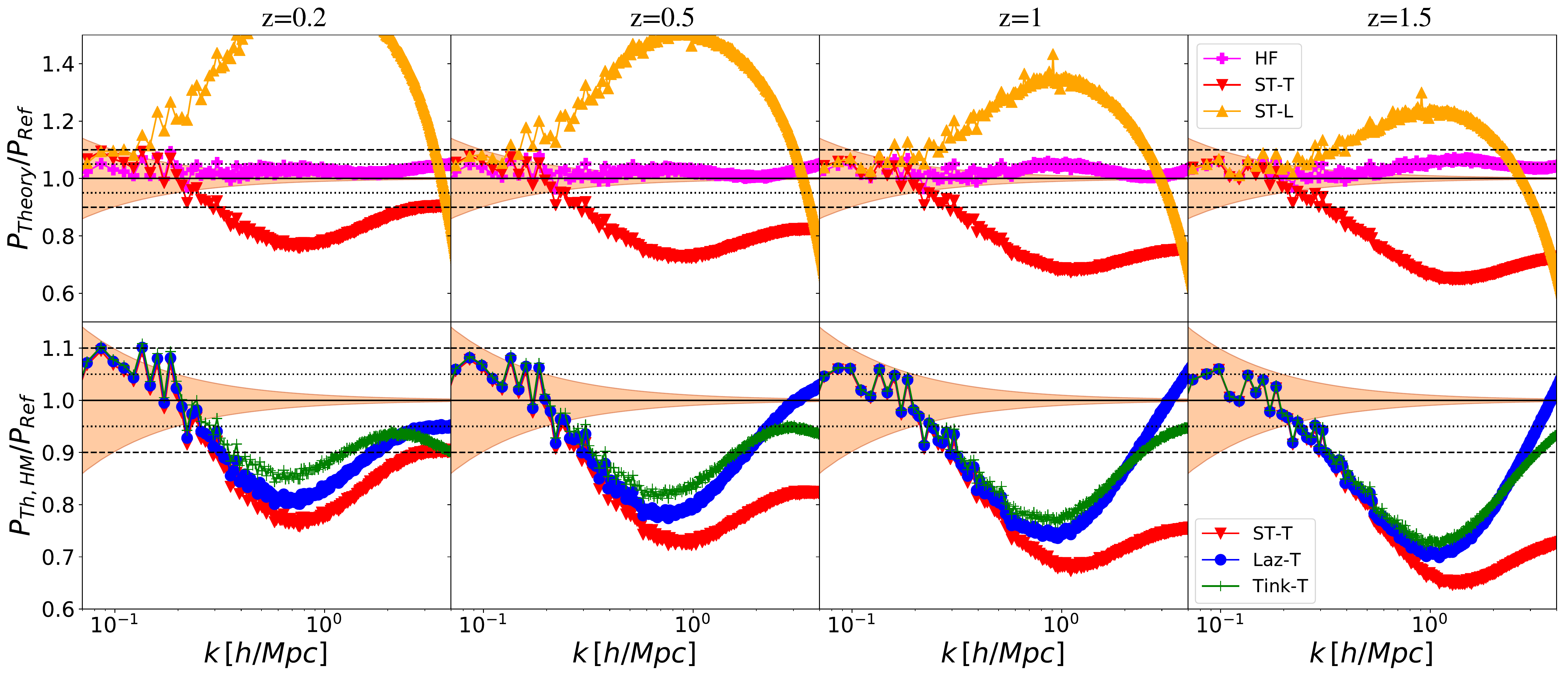}\\
  \caption[CONVERGENCE]{ Comparison of the models for the matter power spectrum with $N$-body simulations in GR, from left to right: $z = 0.2, 0.5, 1$.  We refer the reader to table.~\ref{model_table_1} and table.~\ref{model_table_2} for abbreviations. {\cb \textit{Top:} ratio of various models to the $N$-body measurements:} HF (magenta full pluses) and the ST-T and ST-L halo model prescriptions (red and orange triangles). {\cb \textit{Bottom:} same ratio, concentrating on the halo model prescriptions and varying the halo model ingredients:} ST-T (red triangles), Laz-T (blue circles) and Tink-T (green pluses) halo model prescriptions. The dashed and dotted lines denote $10\%$ and $5\%$ deviations {\cb from the $N-$body measurements} respectively. The error bands are given by twice eq.~\eqref{pserrs}.}
\label{ps_gr}
\end{figure}

We begin by testing our setup with measurements of the power spectrum. Various power spectrum predictions for modified gravity have already been compared against simulations (see~\cite{Cataneo:2018cic} and references therein or~\cite{Lombriser:2014dua} for a review on approaches in chameleon gravity).
In particular, we highlight the reaction-corrected fitting formulae
approach of~\cite{Cataneo:2018cic}, which established a generalised and accurate modelling procedure. We provide these results for reasons of completeness and as a means of checking the consistency of the best halo model ingredients when comparing the bispectrum in section~\ref{sec:bispectrum}. The power spectrum results also provide a good comparison for the levels of accuracy between the state-of-the-art modelling of the bispectrum and power spectrum in modified gravity theories. 
\\ 
\\
The upper panel of figure~\ref{ps_gr} shows the ratio of the theoretical predictions to the $N$-body measurements for $\Lambda$CDM. The lower panel focuses on halo model predictions, testing the different prescriptions for the mass function and virial concentration laid out in section~\ref{sec:halobipo}. {\cb We have split the predictions like this in order to separate comparisons of various modelling procedures and comparisons within the halo model.}
From the upper panels of figure~\ref{ps_gr}, we see that by far the most accurate prescription is halofit (HF), being within $5\%$ accurate up to $k=4h/{\rm Mpc}$. Finally, we note that the ST-T does fairly well at low redshifts whereas ST-L does better at high redshifts, again an expected result given the low redshift divergent behaviour of the loop expansion (see~\cite{Carlson:2009it} for example). Both are worse than the $10\%$ level over scales of $k \geq 0.3h/{\rm Mpc}$ at all redshifts. The lower panels show the  predictions ST-T (red crosses), Laz-T (blue circles) and Tink-T (green triangles). It is clear that the Tinker mass function with the power-law concentration (Tink) does the best overall but still shows deviations of up to $30\%$ at $k\sim1h/{\rm Mpc}$ at high redshifts. We have also checked that the ST-L, Tink-L, Laz-L do worse than ST-T, Tink-T, Laz-T, for $z < 1$ with the two being comparable for $z\geq1$ in the ST case. 
\begin{figure}[htbp!]
\centering
  \includegraphics[width=\textwidth,height=7cm]{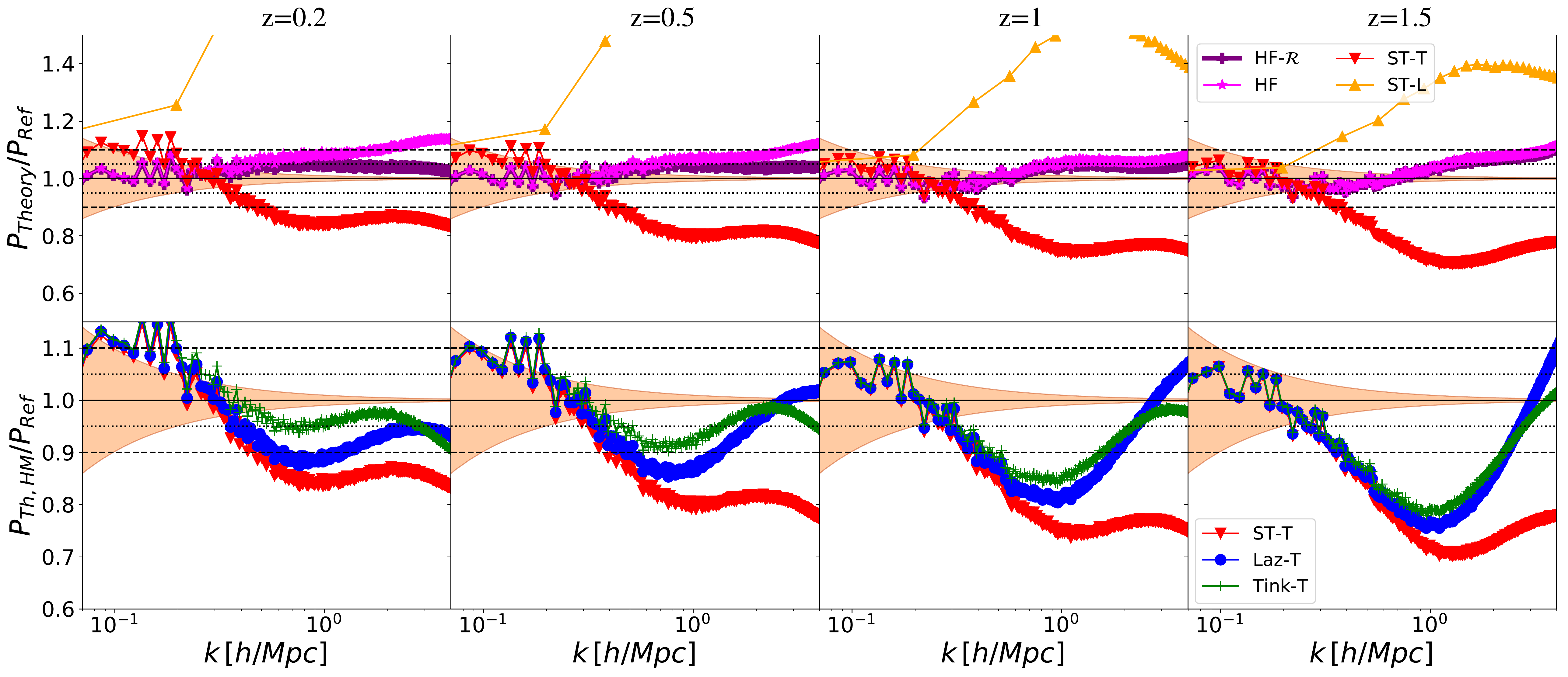}\\
  \caption[CONVERGENCE]{ Same as figure~\ref{ps_gr} but in $f(R)$ gravity with $|f_{\rm R0}| = 10^{-5}$ (F5). We additionally show the HF-$\mathcal{R}$ model (purple pluses) as well as the HF formula (magenta stars).
  }
\label{ps_f5}
\end{figure}
\\ 
\\
 Figure~\ref{ps_f5} shows the power spectrum results for $f(R)$ gravity. These are qualitatively similar to the $\Lambda$CDM results. Again we find that Tink-T is the best prescription when considering the halo model. Here we also consider the halofit formula by simply replacing all linear power spectra  with the linear $f(R)$ power spectrum (HF). This prescription still does better than the pure halo model prescriptions, despite the formula not being fit to $f(R)$ simulations. This prediction can be improved even further by including the `reaction' correction as briefly described at the end of section~\ref{sec:halobipo} and introduced in \cite{Cataneo:2018cic} (HF-$\mathcal{R}$, eq.~\eqref{eq:react}).  We show both the reaction corrected and uncorrected halofit spectra in the upper panels as purple pluses and magenta stars respectively. The HF-$\mathcal{R}$ remains within $\sim 5\%$ of the simulation measurements up to $k=4 \ h/{\rm Mpc}$ for all redshifts considered, whereas the best halo model prescription shows deviations of up to $30\%$ at $z=1.5$, where it performs worst. 
\begin{figure}[htbp!]
\centering
  \includegraphics[width=\textwidth,height=7cm]{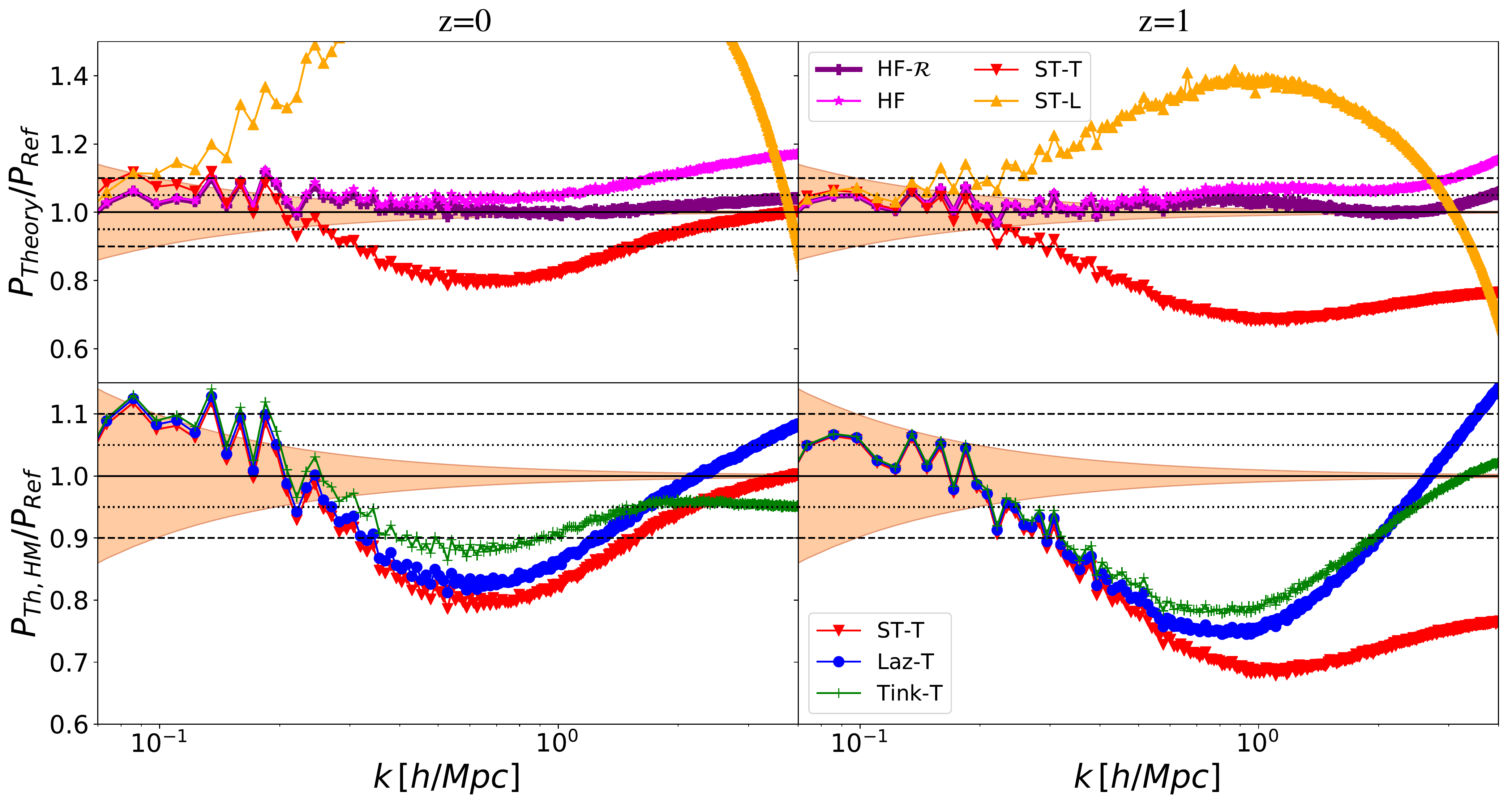}\\
  \caption[CONVERGENCE]{ Same as figure~\ref{ps_gr} but for DGP gravity with $H_0 r_c = 0.5$. In this case we only have $N$-body measurements at $z=0$ (left) and $z=1$ (right). As in figure~\ref{ps_f5}, we additionally show the HF-$\mathcal{R}$ model (purple pluses) as well as the HF formula (magenta stars).}
\label{ps_dgp}
\end{figure}
\\ 
\\
\noindent Figure~\ref{ps_dgp} shows the power spectrum results for DGP. These are again qualitatively similar to the $f(R)$ results with the HF-$\mathcal{R}$ performing the best, maintaining almost percent-level accuracy at all scales and both redshifts (these results are not new, see \cite{Cataneo:2018cic}). Again Tink-T does the best among the pure halo model approaches.
\begin{figure}[htbp!]
\centering
  \includegraphics[width=\textwidth,height=7cm]{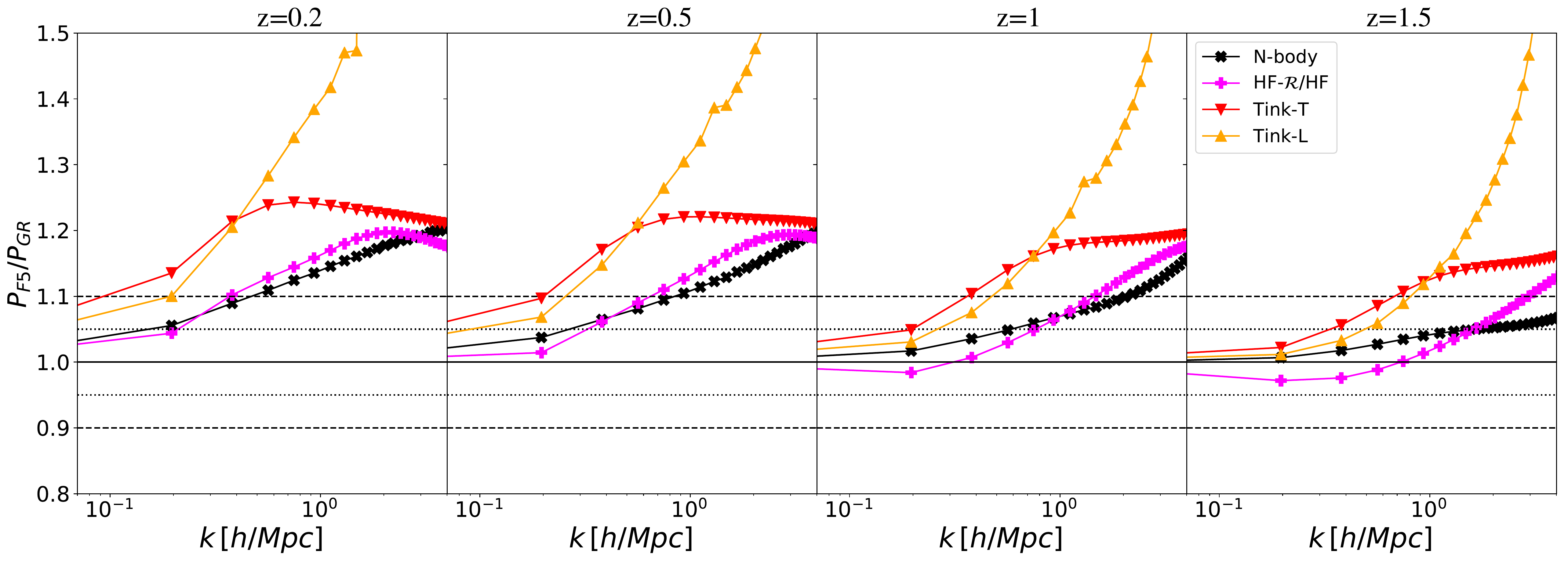}\\
  \caption[CONVERGENCE]{The ratio of the  matter power spectrum in $f(R)$ to the same quantity in $\Lambda$CDM. Simulations are shown as full black crosses, the Tink-T and Tink-L as red down and orange up triangles respectively, and the halofit formulae as magenta full pluses. Regards the fitting formulae, we use the HF-$\mathcal{R}$ for $f(R)$ and HF for GR. The ratio is shown for redshifts $z=0.2,0.5,1$ and $1.5$ from left to right. {\cb  The horizontal dashed and dotted black lines denote $10\%$ and $5\%$ deviation from the GR prediction respectively.} }
\label{ps_f5r}
\end{figure}
\\ 
\\
\noindent Figure~\ref{ps_f5r} shows the ratio of the $f(R)$ spectra to their $\Lambda$CDM counterparts. We see that the fitting formulae (HF and HF-$\mathcal{R}$) model the ratio very well with the Tink-T model performing well at small and large scales but showing $\sim 10\%$ deviations on scales of $0.2h/{\rm Mpc} \leq k \leq 3h/{\rm Mpc}$, where the power spectrum transitions between the 2- and 1- halo terms. Figure~\ref{ps_dgpr} shows the same ratio at $z=0$ and  1 for DGP, where we have normalised the ratio to unity at large scales. Here we find that all matter power spectrum prescriptions model the ratio fairly well. This is somewhat expected as the dominant modification at $z=1$ comes from the overall re-scaling of the linear growth factor. 
\begin{figure}[t]
\centering
  \includegraphics[width=\textwidth,height=7cm]{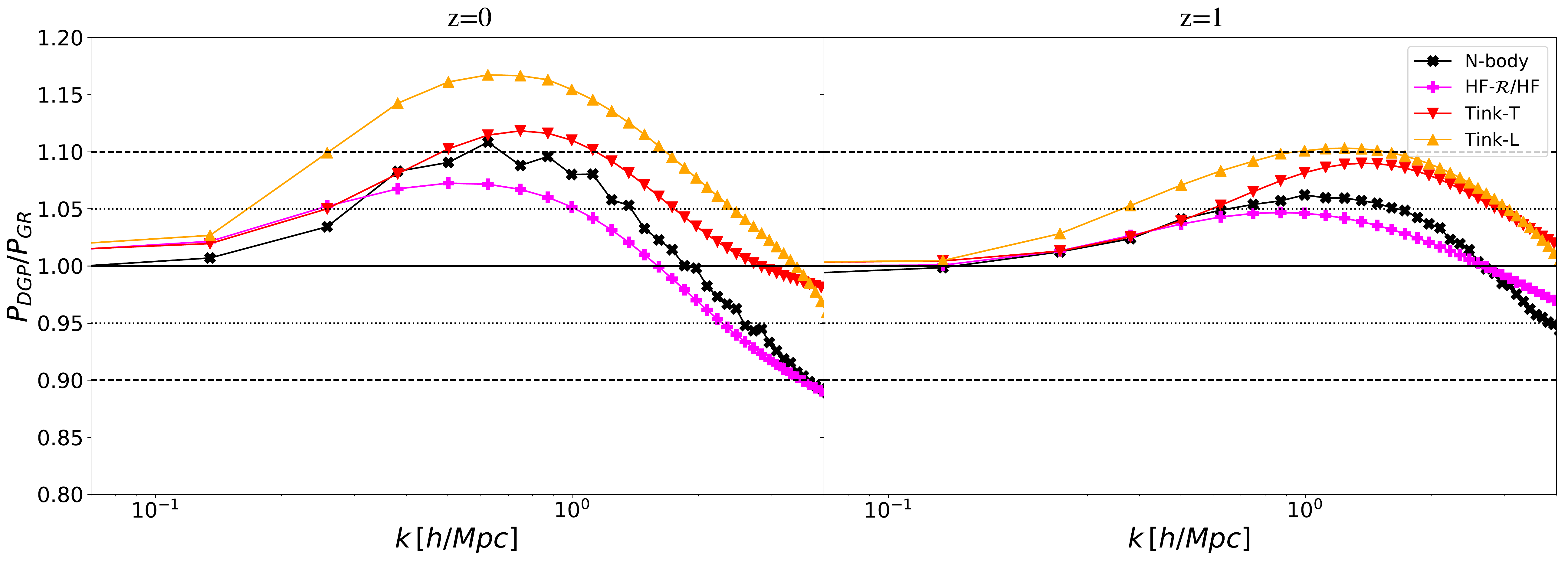}\\
  \caption[CONVERGENCE]{
Same as figure~\ref{ps_f5r} but showing the ratio of DGP gravity to $\Lambda$CDM, and at $z=0$ and $z=1$. Note that we have normalised all DGP spectra to match $\Lambda$CDM at large scales for better viewing. The ratio of the linear spectra are almost a constant over all scales.
}
\label{ps_dgpr}
\end{figure}
\\
\\
Finally, we remark that figure~\ref{ps_f5r} and figure~\ref{ps_dgpr} highlight a merit of the pure halo model: modelling the modification to GR. The Tink-T model does fairly well in both $f(R)$ and DGP in modelling this modification, which can then be used to rescale an accurate prediction for the $\Lambda$CDM power spectra, as done in the more sophisticated reaction approach of \cite{Cataneo:2018cic}. Such accurate predictions for the $\Lambda$CDM power spectrum are readily available in the form of emulators \cite{Knabenhans:2018cng,Heitmann:2015xma}. We note here that such emulators for the bispectrum are not available and it is yet to be seen if the pure halo model approaches can model this modification to GR accurately for the bispectrum. On this note, we move on to our bispectrum results.

\subsection{Bispectrum results}\label{sec:bispectrum}
\begin{figure}[htbp!]
\centering
  \includegraphics[width=\textwidth,height=7cm]{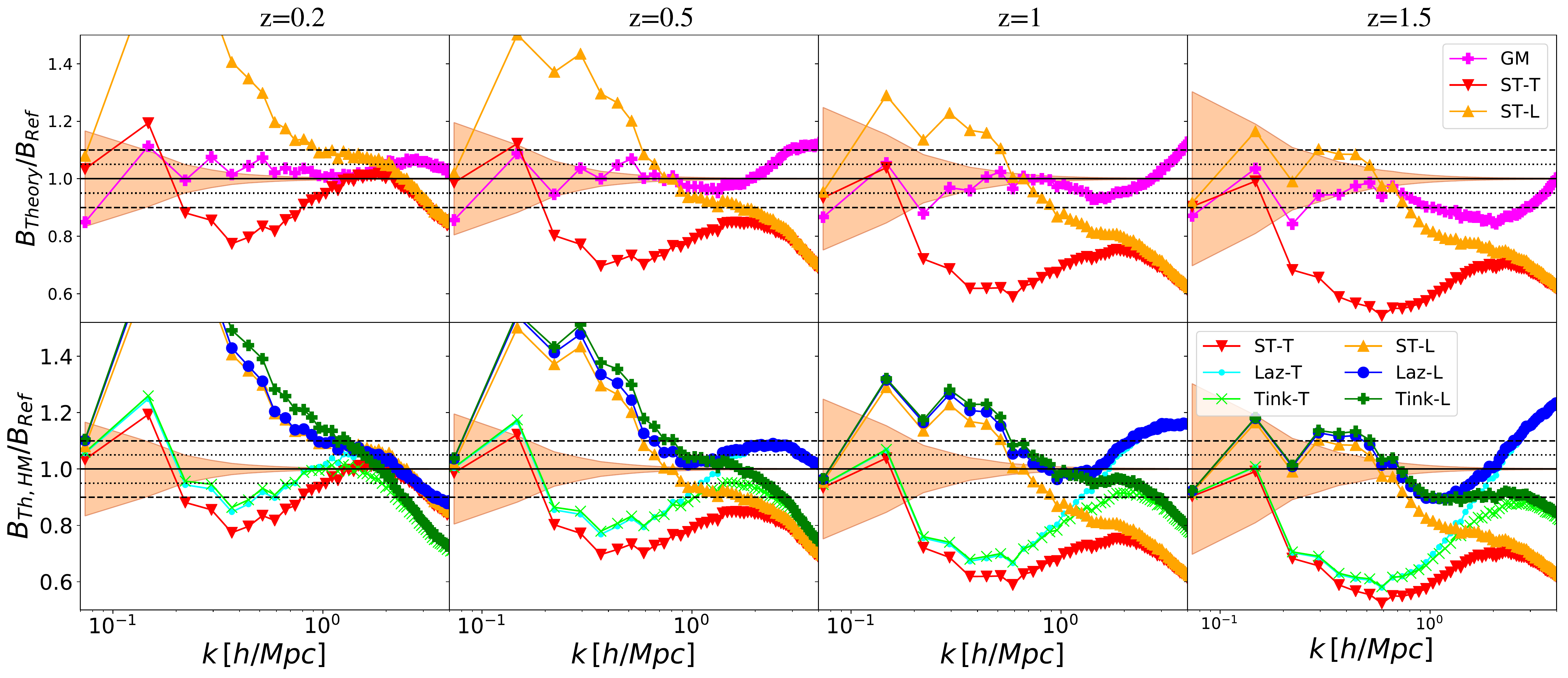}\\
  \caption[CONVERGENCE]{ Comparison of the models of the equilateral matter bispectrum with $N$-body simulations at $z = 0.2, 0.5, 1$ and $1.5$ (left to right) in GR. {\cb \textit{Top:} ratio of various models to the $N$-body measurements:} the GM fitting formula (magenta full pluses) and the ST-T and ST-L halo model predictions (red and  orange triangles). {\cb \textit{Bottom:} same ratio, but concentrating on the} halo model predictions for various choices of mass function and virial concentration as well as using linear or 1-loop bispectra in the 3-halo term. These choices are ST-T/ST-L (red/orange triangles), Laz-T/Laz-L (cyan dots/blue circles) and Tink-T/Tink-L (lime crosses/green pluses). The dashed and dotted lines denote $10\%$ and $5\%$ deviations respectively. The error bands are given by twice eq.~\eqref{bserrs}.}
\label{bs_gr}
\end{figure}

\begin{figure}
\centering
  \includegraphics[width=\textwidth,height=7cm]{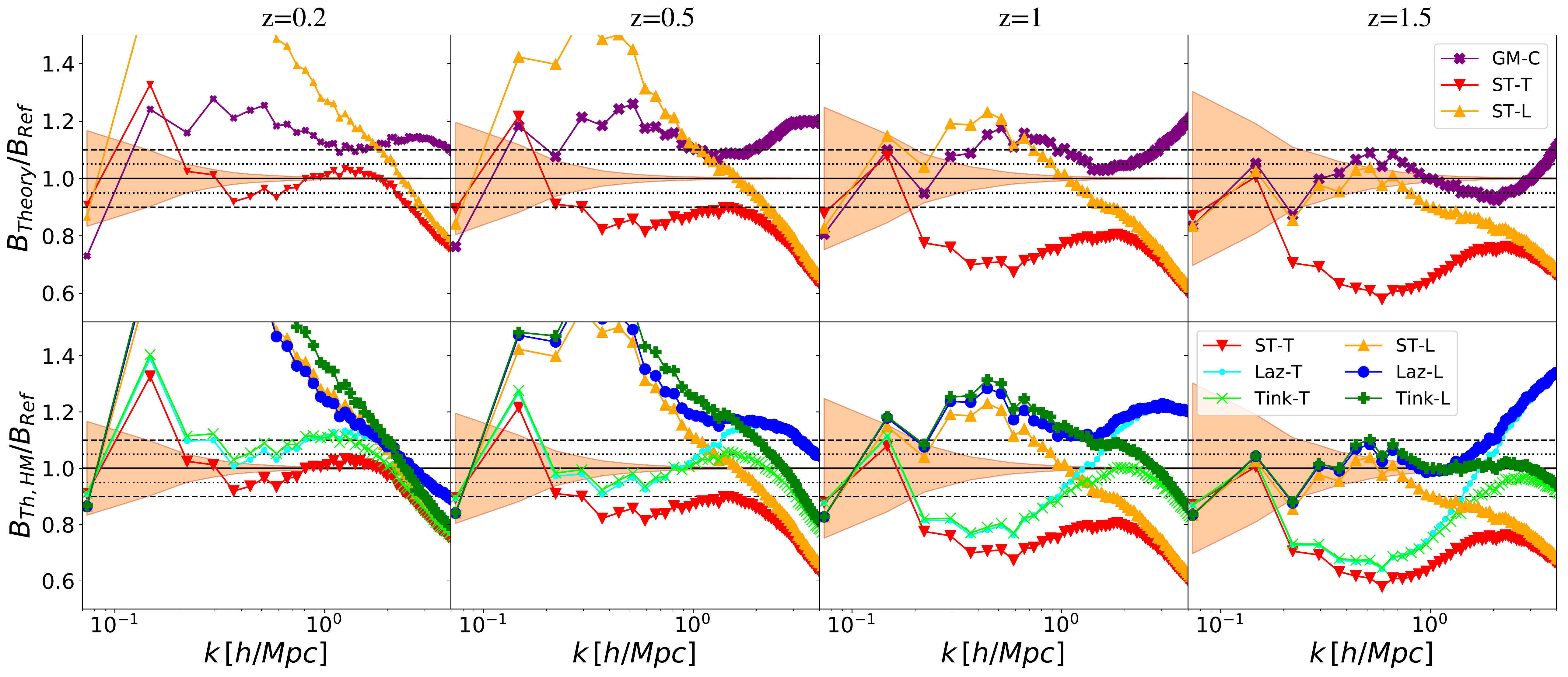}\\
  \caption[CONVERGENCE]{ Same as figure~\ref{bs_gr} but for $f(R)$ gravity with $|f_{\rm R0}| = 10^{-5}$ (F5). Here the purple crosses  show the corrected GM formula (GM-C).}
\label{bs_f5}
\end{figure}

\begin{figure}
\centering
  \includegraphics[width=\textwidth,height=7cm]{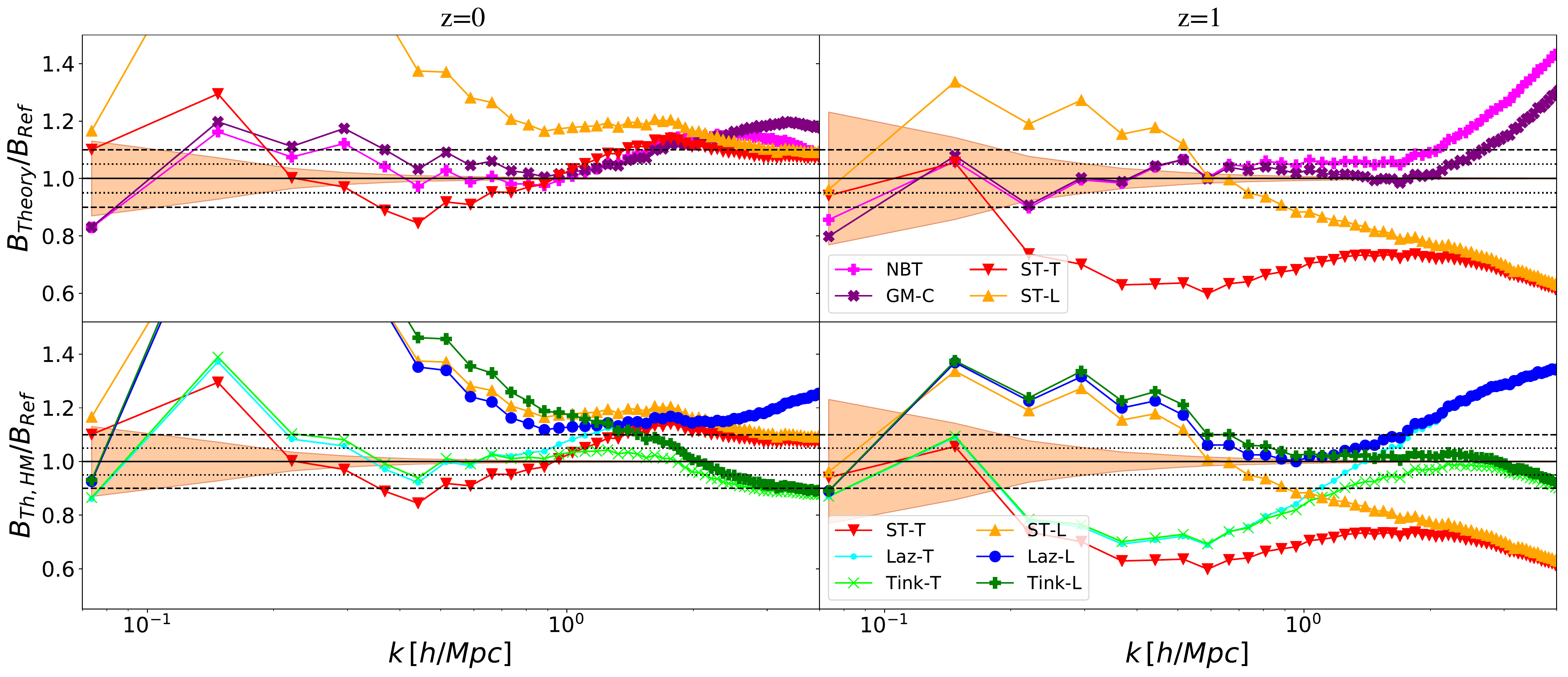}\\
  \caption[CONVERGENCE]{ Same as figure~\ref{bs_f5} but for DGP gravity with $H_0 r_c = 0.5$. Here the magenta pluses show the NBT formula.}
\label{bs_dgp}
\end{figure}

We now present the results for the bispectrum. We only consider the equilateral configuration. This is motivated by \cite{Bose:2018zpk} where it was shown to have the largest signal of modified gravity in the quasi-non-linear regime. This may not be true in the fully non-linear regime, but testing general shapes is computationally difficult in modified gravity and beyond the scope of this initial study. Further, this shape was found to be negligibly affected by binning \cite{Namikawa:2018erh} which we do not consider here. 
\\
\\
The figures in this subsection follow a similar format to those in the previous subsection (see for example figure~\ref{ps_gr}), generally showing the ratios of the theoretical predictions to $N$-body measurements. Upper panels show the fitting formula of \cite{GilMarin:2011ik,Namikawa:2018erh} (GM and NBT, eq.~\eqref{eq:bfit}), the corrected  GM fitting formula (GM-C, eq.~\eqref{eq:correctedgm}, for DGP and $f(R)$  only) and the \cc{ST-T and ST-L} models (recall that these employ the Sheth-Tormen mass function and a power-law virial concentration). Lower panels show the same ratio but the theoretical predictions are for different halo model prescriptions. For the GM-C formula, we  use the Tink-T prescription for the correction factor in eq.~\eqref{eq:correctedgm} as it was shown to  perform the best in power spectrum comparisons.
\\ 
\\
Figure~\ref{bs_gr} shows the results for $\Lambda$CDM. The GM fitting formula (magenta pluses) outperforms all other predictions, staying within $15\%$ of the measurements at all scales and redshifts considered. The ST-T model again does better than the ST-L model at large scales and $z=0.2$, but for $z\geq 0.5$ ST-L outperforms ST-T. However, ST-L still shows $40\%$ deviations at $k=4h/{\rm Mpc}$ and $z\geq 1$. In the lower panels we now show both ST, Tink, Laz-T and ST, Tink, Laz-L predictions for the various halo model prescriptions. As with the power spectrum we find that the Tinker mass function (Tink) outperforms the Sheth-Tormen mass function (ST) but it is unclear which virial concentration does best for the scales and redshifts considered.
\\
\\
At low redshift the 1-loop bispectrum suffers from divergences common to the SPT approach~\cite{Carlson:2009it}. Motivated by this, we have also checked for improvement of the ST-L prediction by introducing the resummation prescription of~\cite{Bernardeau:2008fa,Taruya:2012ut} \cc{in the 1-loop calculation within the 3-halo term}. We find that at $z\geq 0.5$ it performs worse than the unresummed version, and is comparable in accuracy to the ST-T prediction at $z=0.2$.
\\
\\
\noindent Figure~\ref{bs_f5} shows the results for $f(R)$ gravity. Qualitatively, the results are the same as those for $\Lambda$CDM except for the improvement of the power-law (Tink) over the Bolshoi concentration (Laz) in the bottom panels. This may be understood in part through the exclusion of any scale dependent growth\footnote{Recall that we take the small $k$ limit of $F_1(k;a)$, which in $f(R)$ gravity is simply the GR growth factor.} (eq.~\eqref{eq:bol}) or mass dependence of $M_\star$. Furthermore, the time evolution is completely fit to that of GR. The difference between Laz and Tink prescriptions indicates the importance of a general concentration relation for theories beyond GR. We find that the GM-C model,  which multiplies  the $\Lambda$CDM non-linear GM formula  with a ratio of Tink-T models (see  eq.~\eqref{eq:correctedgm}),  does the best overall, staying  within $20\%$ accurate for all  scales  and redshifts considered.  This  is  comparable to the GM formula in the $\Lambda$CDM comparisons. 
\\
\\
\noindent Figure~\ref{bs_dgp} shows the results for DGP gravity. The NBT formula, eq.~\eqref{eq:bfit}, performs the best, staying mostly within $\sim 10\%$ at $z=0$ and diverging significantly beyond $10\%$ at $k>2h/{\rm Mpc}$ at $z=1$.  We have also checked the performance of the GM formula, where we set $\kappa(a)=\lambda(a) =1$ in eq.~\eqref{eq:bfit} but keep the linear growth factor of DGP. We find that the improvement provided by $\lambda_{\rm DGP}(a)$ in eq.~\eqref{fittingbh} is negligible for all scales and redshifts considered. Further, we  also show the GM-C formula and find that it does slightly better over all scales at $z=1$, and equally well at $z=0$ than the NBT formula. From the halo model prescriptions, the Tink-L does very well at $z=1$, whereas the Tink-T does very well at $z=0$. Again, \cc{the poor behaviour of the 1-loop bispectrum at low redshift \cite{Carlson:2009it,Bose:2018zpk} is probably to blame for this trend.}
\\
\\
As for the $\Lambda$CDM case, we have also checked whether using the resummed 1-loop bispectrum improves the predictions for ST-L at all redshifts but find that it performs comparably to the ST-T case at $z=0$ and does significantly worse for $z\geq 0.5$.
\begin{figure}
\centering
  \includegraphics[width=\textwidth,height=7cm]{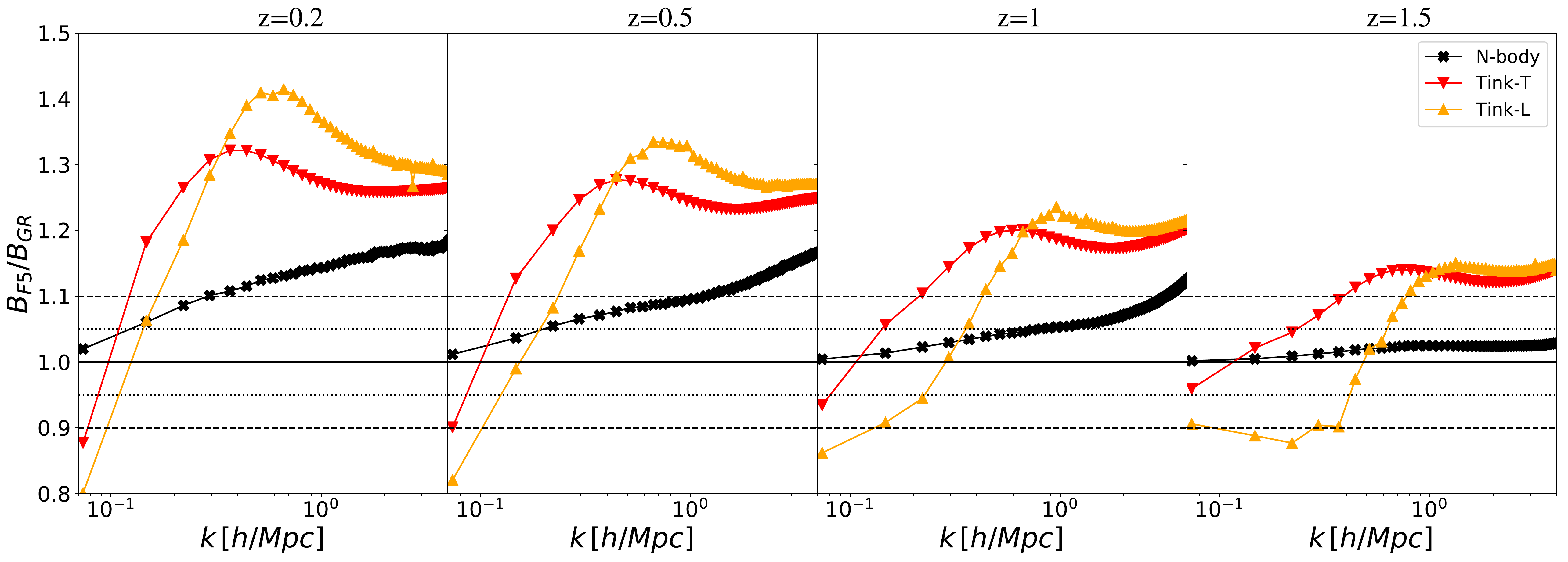}\\
  \caption[CONVERGENCE]{The ratio of the equilateral matter bispectrum in $f(R)$ gravity to the same quantity in $\Lambda$CDM. Simulations are shown as full black crosses, the Tink-T and Tink-L models are as red down and orange up triangles respectively. The ratio is shown for redshifts $z=0.2,0.5,1$ and $1.5$ from left to right. }
\label{bs_f5r}
\end{figure}
\begin{figure}
\centering
  \includegraphics[width=\textwidth,height=7cm]{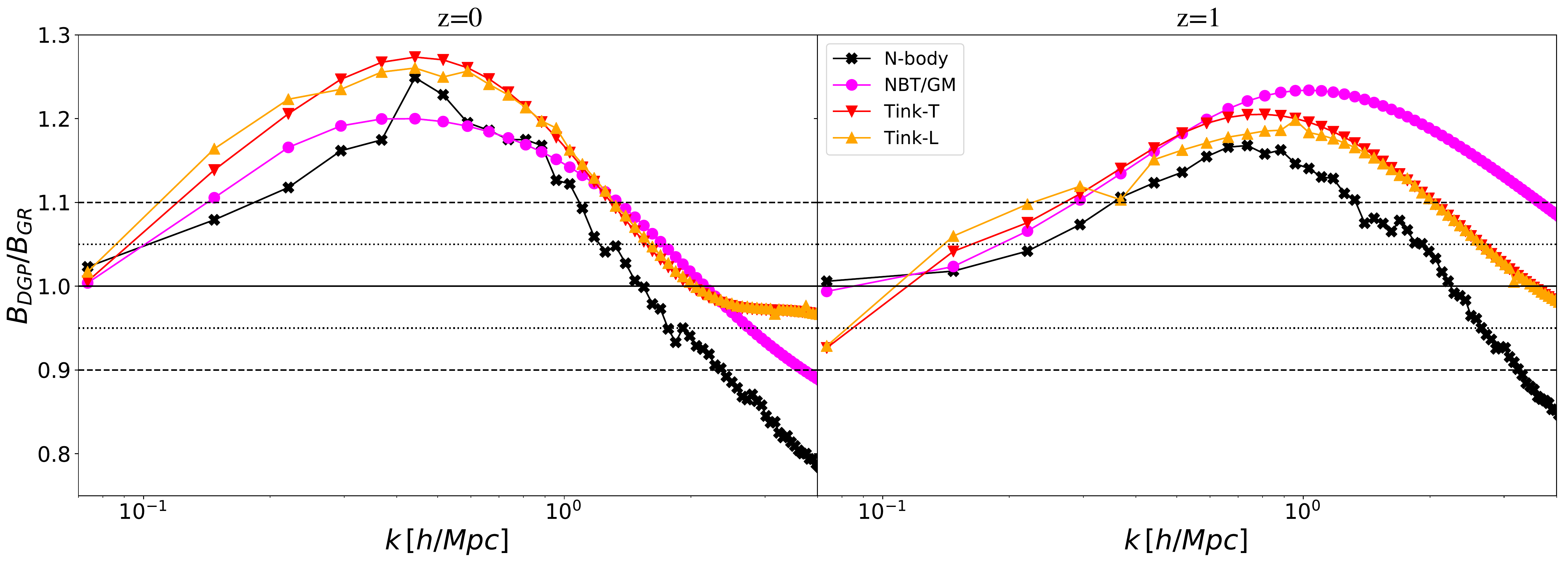}\\
  \caption[CONVERGENCE]{Same as figure~\ref{bs_f5r} but showing the ratio of DGP gravity to $\Lambda$CDM  and at $z=0$ and $z=1$. We also show here the ratio of the fitting formulae as magenta circles. Note that we have normalised all DGP spectra to match $\Lambda$CDM at large scales for better viewing.
  }
\label{bs_dgpr}
\end{figure}
\\
\\
\noindent Figure~\ref{bs_f5r} shows the ratio of the $f(R)$ to $\Lambda$CDM  matter bispectrum predictions. In this way, inaccuracies in the spectrum modelling are  reduced and we have a better measure of the accuracy of the gravitational modelling (note that the simulations have the same initial seeds so cosmic variance is essentially cancelled by taking the ratio). Here we see that the Tink-T and Tink-L models do fairly well in modelling the gravitational effects in $f(R)$ gravity, at all redshifts. This fact is reflected in the accuracy of the GM-C formula shown in purple in figure~\ref{bs_f5}. 
\\
\\
{\cb The red Tink-T curve at $z=0.2$ in figure~\ref{bs_f5r} is our most readily comparable result to the results of \cite{Brax:2012sy}. In particular, they show an unscreened version of the halo model predictions in their figure 19. Both our results show a peak deviation in the quasi non-linear regime, $k\leq 0.3h/{\rm Mpc}$. Interestingly, we find that at non-linear scales, the deviation from $\Lambda$CDM compared to the results in \cite{Brax:2012sy} is enhanced. This is somewhat counter-intuitive since a main difference in the results is the inclusion of screening effects within spherical collapse in this work, which should work to suppress deviations. Despite this, our results are consistent with the trend of the N-body measurements. This suggests that the scales $k\leq 3h/{\rm Mpc}$ are still well above the screened regime at the redshifts considered, making non-linear modifications to gravity relevant at these scales.}
\\
\\
Figure~\ref{bs_dgpr} shows the ratio of the DGP predictions to $\Lambda$CDM at $z=0$ and $1$. We normalise the curves to the ratio of the tree level (see eq.~\eqref{tree2}) predictions at large scales since within DGP the linear growth factor sees a constant enhancement with respect to $\Lambda$CDM\footnote{ \cc{The tree level  ratio was checked to be almost constant over all scales considered, indicating second-order non-linearities introduced by screening effects are negligible.}}. Importantly, we find here that the halo model performs better than the fitting formulae at modelling the ratio, and that there is little difference (at least at $z=1$) between Tink-T and Tink-L at scales above $k\sim 1h/{\rm Mpc}$. 
\\
\\
 The GM-C formula does relatively well for both DGP and $f(R)$, reflected in the accuracy of the Tink-T curves in figure~\ref{bs_f5r} and figure~\ref{bs_dgpr}. These plots also suggest that an improved correction, such as the reaction method \cite{Cataneo:2018cic}, may be a promising approach to improve the modelling of the bispectrum. The reaction method relies on the halo model to give the correction to non-linear gravitational dynamics in modified theories, and we find here that the halo model seems to do this sufficiently well for the bispectrum.
 \\
 \\
 \cc{Finally, we make a brief comment on baryonic effects. Recently these have been studied thoroughly for GR in \cite{Foreman:2019ahr}, where they see the baryonic physics becoming important at small scales, with up to a $20\%$ effect at $0\leq z \leq 2$ and $k\approx 4h/{\rm Mpc}$. We find inaccuracies in the matter bispectrum that are comparable to the baryonic effects in both modified gravity theories considered here. This motivates further work on the matter statistics before considering modelling baryons. Further, for DGP the peak of the signal occurs at scales where baryonic physics is not important (see figure~\ref{bs_dgpr}). While it remains unclear if this is a Vainshtein specific feature, it does indicate that promising constraints can be achieved on scales where baryonic physics can be safely neglected. }


\section{Lensing bispectrum: impact of non-linearities and inaccuracies}\label{sec:lenscomp}
In this section we compare various models for the matter bispectrum at the level of the lensing convergence (see  eq.~\eqref{eq:lensbs}). This gives us an indication of the impact of inaccuracies in modelling the matter bispectrum on the lensing convergence bispectrum. \cc{We also include the tree level predictions for comparisons (see eq.~\eqref{tree2}} 
\\
\\
\begin{figure}[htbp!]
\centering
  \includegraphics[width=\textwidth,height=7cm]{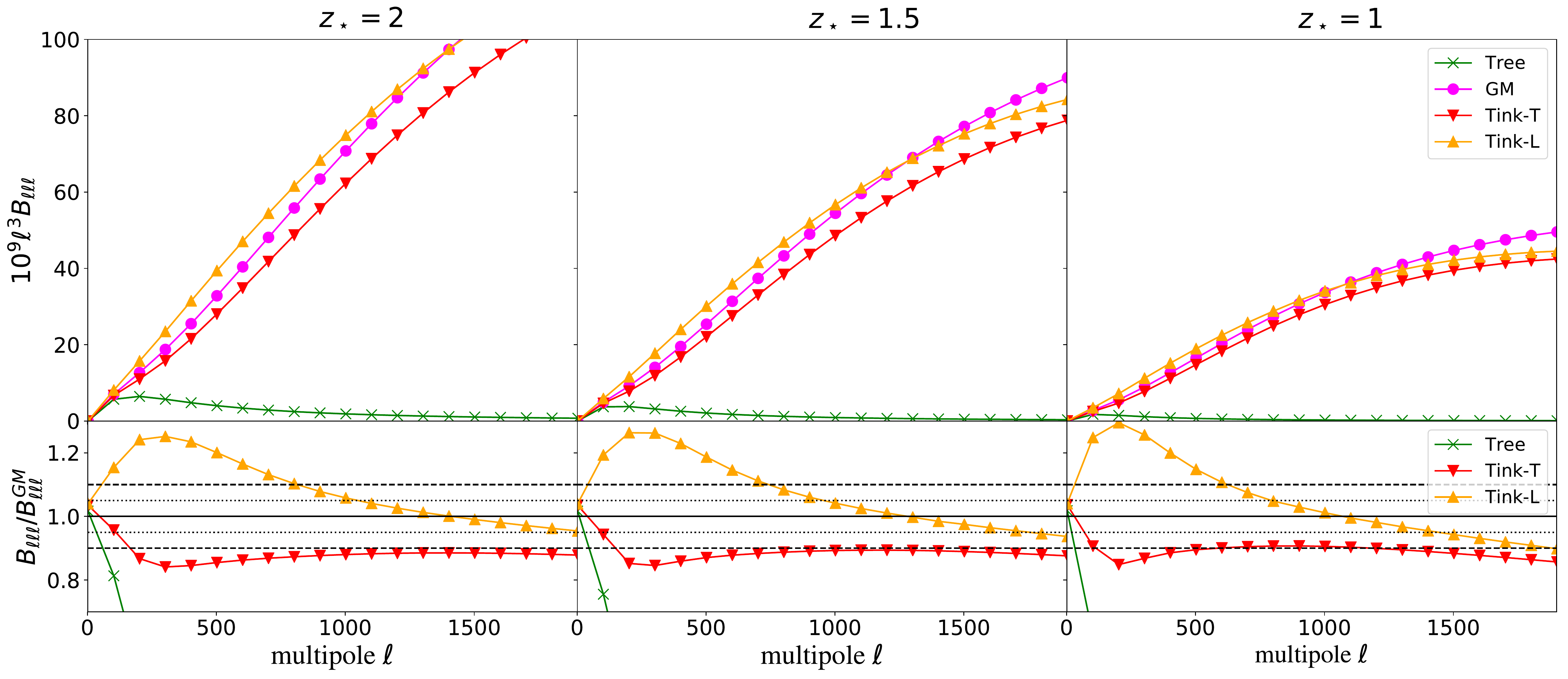}\\
  \caption[CONVERGENCE]{Upper panels show the equilateral bispectrum lensing convergence spectrum in $\Lambda$CDM for four prescriptions for the matter bispectrum: tree level (green crosses), GM fitting formula (magenta circles), Tink-T (red triangles) and Tink-L (orange down triangles) as a function of multipole $
  \ell$. The bottom panels plot the ratio of the models with the GM formula. We include the results for three different source redshifts $z_\star=2$ (left), $z_\star=1.5$ (middle) and $z_\star=1$ (right). }
\label{bl_gr}
\end{figure}

\begin{figure}[htbp!]
\centering
  \includegraphics[width=\textwidth,height=7cm]{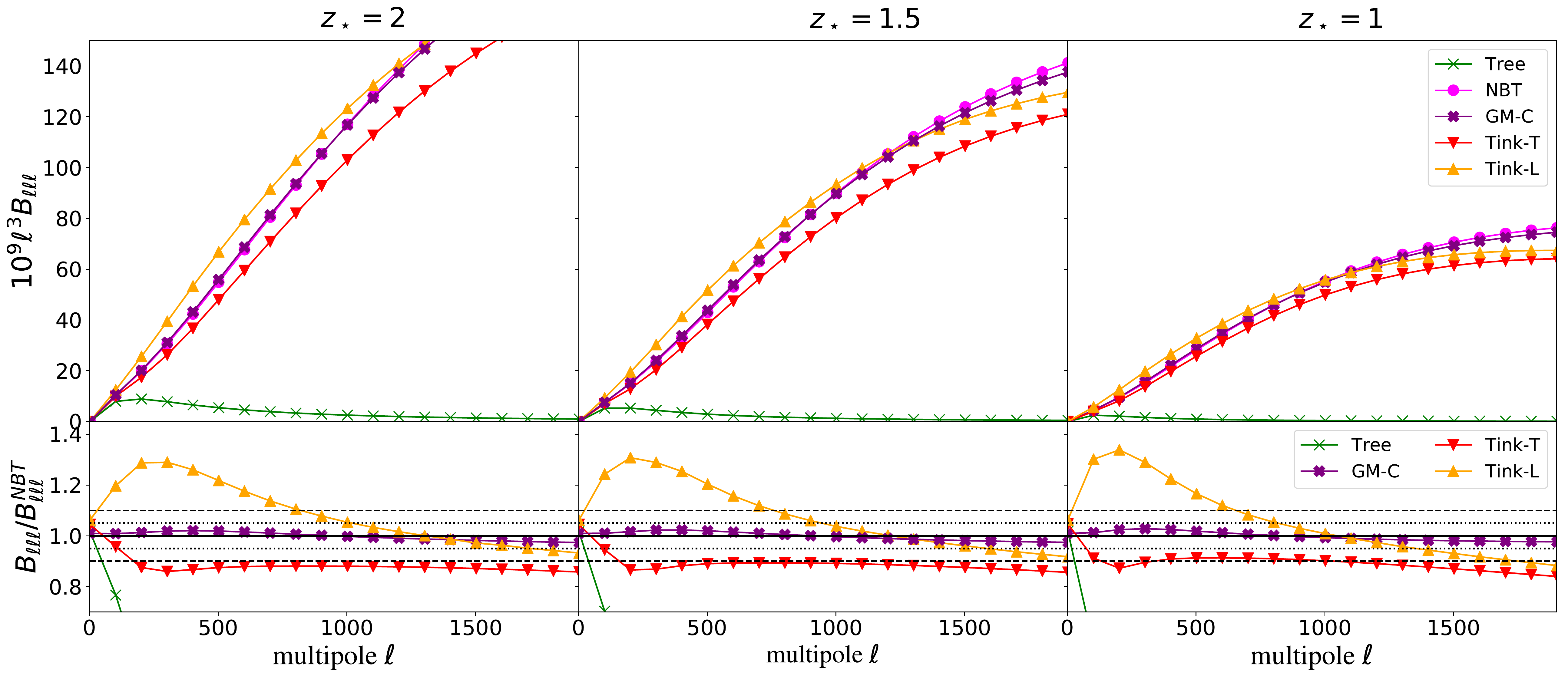}\\
  \caption[CONVERGENCE]{Same as figure~\ref{bl_gr} but for DGP gravity. The magenta circles are now the NBT formula and the purple full crosses is the GM-C prediction.} 
\label{bl_dgp}
\end{figure}

\begin{figure}[htbp!]
\centering
  \includegraphics[width=\textwidth,height=8cm]{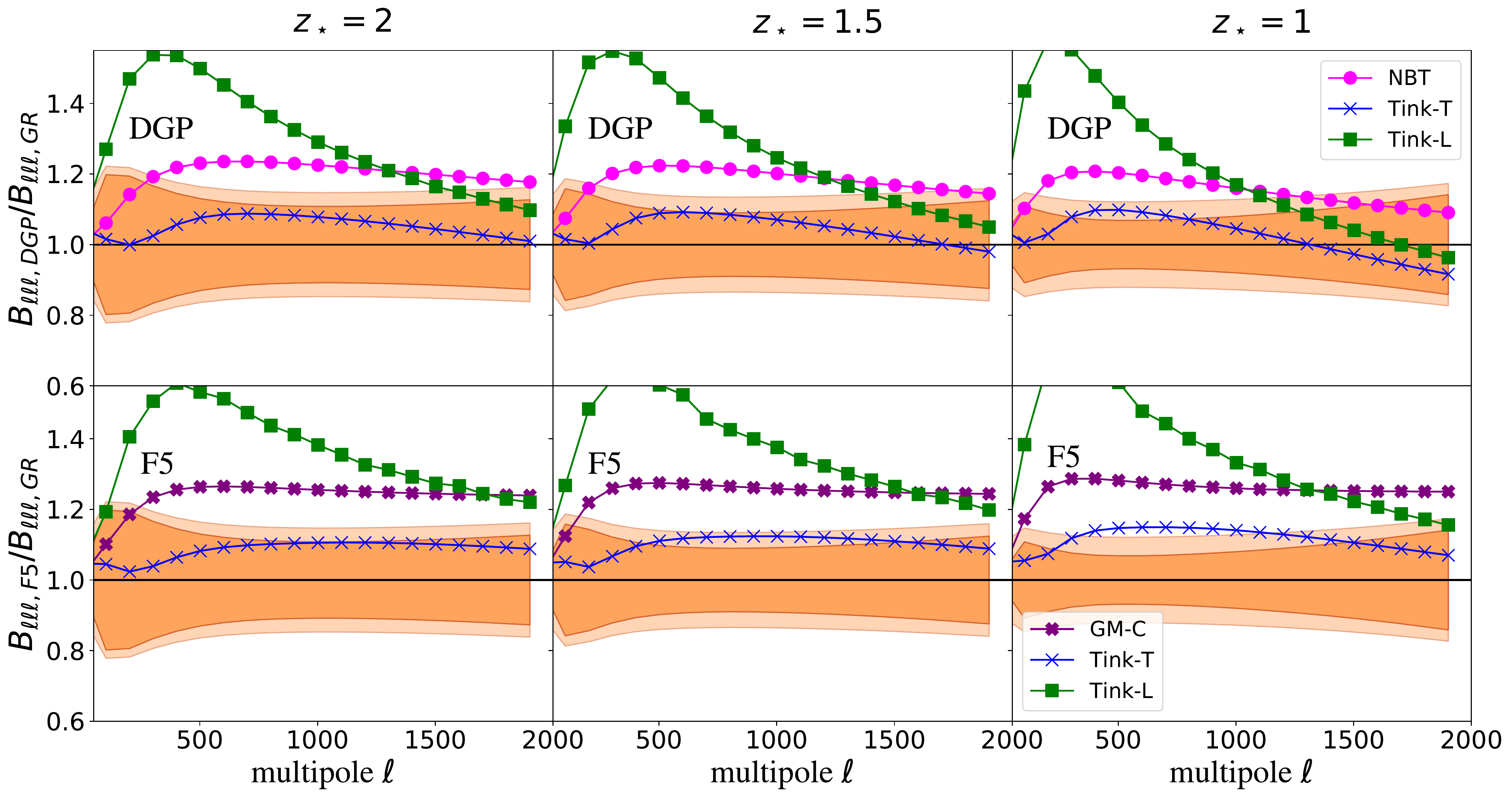}\\
  \caption[CONVERGENCE]{The ratio of the equilateral lensing convergence  bispectrum in DGP (top panels) and $f(R)$ (bottom panels) to the same quantity in $\Lambda$CDM. The $\Lambda$CDM values are computed using the GM fitting formula for the matter bispectrum. For the modified gravity models we show the Tink-T (blue crosses), Tink-L (green squares), NBT fitting formula (magenta circles, DGP only) and GM-C formula (purple full crosses, $f(R)$ only). The ratio is given for source redshifts $z_\star=2,1.5$ and $1$  from left to right. The orange error bands are the $2\sigma_b$ errors, where $\sigma_b$ is given by eq.~\eqref{eq:bl_error}, while the beige error bands also include a $10\%$ modelling error, added in quadrature. For the DGP plots we normalise the ratio to unity at $\ell=2$. } 
\label{bl_all}
\end{figure}

\noindent In figure~\ref{bl_gr} we consider $\Lambda$CDM for three different source redshifts, $z_\star=2$ (left), $z_\star=1.5$ (middle) and $z_\star=1$ (right). We show the equilateral configuration for the lensing bispectrum for four prescriptions for $B_{\delta}$: tree level, the GM fitting formula, the Tink-T and Tink-L halo models. In the absence of ray tracing simulations and given that the GM formula is by far the most accurate model for the matter bispectrum we have, it sets a good benchmark with which to compare the other prescriptions. Testing the GM formula against high quality simulations for different lensing bispectrum shapes is the focus of \cite{Munshi:2019csw}, where it has been shown to be very accurate for $\ell\leq2048$ for the equilateral case at the source redshifts considered here.
\\
\\
The ratio of the GM lensing bispectrum with the other prescriptions is given in the bottom panels of figure~\ref{bl_gr}. The tree level diverges at extremely small multipoles whereas Tink-T remains at an almost constant offset of $\sim 10\%$ for all source redshifts. Tink-L on the other hand does badly (up to $\sim30\%$ deviation) at small multipoles but for $\ell > 750$ remains within $10\%$. 
\\
\\ 
In figure~\ref{bl_dgp} we plot the results for DGP, which are again qualitatively the same as the $\Lambda$CDM results. Taking the NBT formula as our benchmark in accuracy, we again see a constant $10\%$ deviation of Tink-T for $\ell>150$ while Tink-L becomes more accurate than this at $\ell \geq 750$ at source redshift $z_\star=2$ and slightly lower at $z_\star=1$. We see that the NBT and GM-C models are within $5\%$ of each other at all multipoles considered reflecting their similarity at the matter bispectrum level. We do not show the pure $f(R)$ spectra as we do not have a significantly accurate prescription for the $f(R)$ matter bispectrum.
\\
\\
\\
\noindent  Furthermore, we inspect the ratios of the modified lensing spectra to those of $\Lambda$CDM. We will also consider errors coming from the lensing bispectrum variance, specifically from the Gaussian contribution~\cite{Kayo:2013aha,Rizzato:2018whp}. For the equilateral shape, these are given by 
\begin{equation}
\sigma_{b}^{\rm equilateral}(\ell) = \sqrt{6 \frac{[C_\ell + \sigma_\epsilon^2/\bar{n}]^3}{N_{\ell \ell \ell } \ \Omega_s (\Delta \ell)^3}},  \label{eq:bl_error}
\end{equation}
with $\Omega_s = 4\pi f_{\rm sky}$, $f_{\rm sky}$ being the sky coverage of the survey. $C_\ell$ is the angular power spectrum, $\sigma_\epsilon$ denotes the shape noise parameter, $\bar{n}$ represents the projected number density of source galaxies per steradians,\footnote{\cc{This will generally depend on survey depth.}} $\Delta \ell$ is the bin width and $N_{\ell \ell \ell}$ is given by~\cite{Lacasa:2011ej}
\begin{equation}
    N_{\ell_1 \ell_2 \ell_3} = \frac{(2\ell_1  + 1) ( 2 \ell_2 + 1 ) ( 2 \ell_3 +1)}{4\pi} \begin{pmatrix} 
\ell_1 & \ell_2 & \ell_3 \\
0 & 0 & 0
\end{pmatrix}^2,
\end{equation}
where for the Wigner-3j symbol we adopt the Stirling approximation (see Appendix~A of~\cite{Takada:2003ef}). We take the central value of the bin when computing $N_{\ell \ell \ell}$. Further, it is assumed that $N_{\ell_1 \ell_2 \ell_3}$ varies slowly within the bin width, so that the binned number of triplets $\sum_{\ell_i \in \mathrm{bin}_i} N_{\ell_1 \ell_2 \ell_3}$ can be approximated with $N_{\ell \ell \ell} (\Delta\ell)^3$.
\\ 
\\ 
To give an indication of the power of the next generation of surveys to detect signals of modified gravity we choose parameters representative of \textit{Euclid}: $\bar{n} = 30 {\rm gal}\cdot{\rm arcmin}^{-2}$, $f_{\rm sky} = 0.36$ and $\sigma_\epsilon = 0.3$~\cite{Laureijs:2011gra,Amendola:2012ys}.
We take a bin width of $\Delta \ell = 100$. Again we quote the $2\sigma$ errors in our plots. 
\\
\\
In figure~\ref{bl_all} the ratio of the equilateral lensing bispectrum of  DGP (top panels)  and  $f(R)$ (bottom panels) to that in GR are given. Plotted are these ratios for halo model prescriptions for the matter bispectrum (Tink-T as blue crosses and Tink-L as green squares) for three different source redshifts, $z_\star=2,1.5$ and $1$. The GR prediction is given using the GM fitting formula for the matter bispectrum.  We also show the NBT formula for DGP as magenta circles and  the  GM-C formula for  $f(R)$ as purple crosses. We do not show the GM-C formula for DGP, which is almost the same as the NBT formula (see figure~\ref{bl_dgp}) to avoid filling the plot unnecessarily. For the DGP plots we normalise the ratio to unity at $\ell=2$. Finally, also shown are the $2\sigma$ errors as given by eq.~\eqref{eq:bl_error} as an orange band. The beige band includes an additional $10\%$ modelling error motivated by the Tink-T inaccuracy indicated in figures~\ref{bl_gr} and \ref{bl_dgp}.
\\
\\
\noindent In figure~\ref{bl_gr} and figure~\ref{bl_dgp}, we found the Tink-T model to be $\sim 10\%$ accurate at most multipoles, while the Tink-L model was more accurate than this at multipoles $\ell \gtrsim750$ for GR and DGP. The signal of modification in both the $f(R)$ and DGP models is well above the bispectrum variance below scales when the Tink-L modelling is most accurate, but Tink-T predictions lie within or close to the error bars for most source redshifts and multipoles. In the DGP case, the fitting formulae ratio shows an intermediate signal, larger than Tink-T but smaller than Tink-L. Despite this, we note that in the DGP case the linear growth factor is enhanced but degenerate with $\sigma_8$, and so most of these enhancements can effectively be absorbed through a rescaling of $\sigma_8$.  The GM-C formula in the  $f(R)$ case remains above both sets of error bands at almost all scales. 
\\
\\
Finally, figure~\ref{bl_all} seems to indicate that for $f(R)$ gravity, low source redshift is slightly preferable in testing this  model, with  a larger deviation from GR being exhibited.  For DGP, Tink-T and Tink-L predict more of a signal at low source redshift whereas the NBT formula predicts a larger signal at high source redshift, although at high redshifts the NBT formula becomes less accurate (see right panel of figure~\ref{bs_dgp}). 
\\
\\
Given the lack of accuracy in the matter bispectrum theoretical prescriptions, we refrain from making any definitive recommendations for a particular choice of model. However, we note that the predicted strength of a modified gravity signal in figure~\ref{bl_all} is strongly dependent on the choice of  model, indicating that modified gravity constraints from the lensing bispectrum would not be robust to the theoretical error introduced by the models that we have explored here. In particular, for  $f(R)$ gravity at $z_\star=1$,  we note  the GM-C and Tink-T predictions lie outside and inside the error bands respectively, whereas in figure~\ref{bs_f5} we see that these models differ only slightly in accuracy at redshifts below $z=1.5$. \cc{We do  note that the level of modification in the gravitational models considered is higher or at the current constraints. Despite this, the wildly different predictions from Tink-T and Tink-L in figure~\ref{bl_all} suggest that more accurate modelling  is needed for future lensing  bispectrum  pipelines. \emph{This is one of the main results of this work.}}


\section{Summary}\label{sec:summary}
We have assessed various theoretical modelling approaches for the matter bispectrum in the non-linear regime of structure formation for some extensions to $\Lambda$CDM, essentially extending the work done in \cite{Bose:2018zpk} to small scales and to a lesser extent, extend the work of \cite{Brax:2012sy} to validation against simulations. In particular we consider the chameleon screened Hu-Sawicki form of $f(R)$ gravity and the Vainshtein screened DGP braneworld model. We compare predictions from the halo model and fitting formulae against measurements from full $N$-body simulations at redshifts relevant for upcoming observations. Furthermore, we test the performances of the Sheth-Tormen and Tinker halo mass functions as well as a power-law virial concentration and the `Bolshoi' simulation concentration  in modelling the two and three point statistics for the considered gravity models.
In our comparisons we only consider the equilateral configuration based on the conclusions of~\cite{Bose:2018zpk} which show that this configuration is sensitive to modifications of gravity in the quasi non-linear regime. We summarise the scales at which the various models for the matter bispectrum remain  within $10\%$ of the $N$-body measurements in table~\ref{result_bstab_total}. \cc{ We choose $10\%$ as a basis as this is roughly the order of $2\sigma_b$ given in eq.~\eqref{eq:bl_error} for stage IV survey-like specifications. }
\\ 
\\
\cc{We  summarise the main results of this work in bullet format before a full discussion.
\begin{enumerate}
    \item 
    All tested prescriptions for the matter bispectrum which are applicable to general theories beyond $\Lambda$CDM are unlikely to be accurate enough for SIV lensing surveys. 
    \item 
    For halo model predictions, the use of a Tinker mass function combined with a power law virial concentration seems to be the best prescription in terms of accuracy for general modifications to GR. 
    \item 
    We propose a halo-model-corrected formula (eq.~\eqref{eq:correctedgm}) based on the $\Lambda$CDM fitting formula of \cite{GilMarin:2011ik} which is both theoretically general and outperforms other prescriptions considered here. 
    \item
    For the models considered, the signal of modified gravity in both the matter and lensing convergence equilateral bispectrum is significant (compared to SIV-like errors) at intermediate scales ($k\leq 1h/{\rm Mpc}$ or $\ell \leq 1000$). This suggests the lensing bispectrum may be a promising means of constraining such theories with the next generation of surveys. 
\end{enumerate}
}
\noindent Within the pure halo models we find that the Tinker mass function combined with a power-law virial concentration relation  performs the best overall, for both power and bispectra and for all gravity models. This provides a good indication of consistency of the halo model itself as well as the robustness of these ingredients when changing the law of gravity. The Bolshoi concentration does comparatively well but only for GR, reflecting the fact that this form is finely tuned to $\Lambda$CDM simulations. The halo model still shows up to $40\%$ disagreement with the simulations for the bispectrum for particular regimes of structure formation and redshifts. \cc{We find ST, Tink, Laz-T all do well at low redshifts while ST, Tink, Laz-L \footnote{\cc{See sec.~\ref{sec:halobipo} for acronym descriptions}.} perform better at high redshifts.} This behaviour  reflects the badly controlled loop expansion of SPT at low redshift. Improvements to the halo model have been made for GR (see  \cite{Lazanu:2015rta} for a comprehensive analysis), but all of these fail in some regime of structure formation or cosmic time, particularly in the transition regime between the 1-halo and 3-halo  terms. Furthermore, some of these improvements rely on higher-order perturbation theory computations \cite{Valageas:2010yw,Valageas:2011up}. The 1-loop bispectrum in modified gravity is very time consuming to compute \cite{Bose:2018zpk} and is not well suited for survey analysis pipelines which will require thousands of bispectrum computations. On the other hand, the Tink-T model has been shown to provide a reasonable fit  to the ratio  of modified gravity to $\Lambda$CDM bispectrum measurements, which can be used to correct $\Lambda$CDM specific non-linear bispectrum prescriptions.
\\
\\
We find that the GM and NBT fitting formulae provide the most accurate  descriptions of the matter bispectrum. These rely heavily on simulations and so do not offer much in the way of generality. In fact, for $f(R)$ gravity there is no such fitting formula for the bispectrum. Instead, we have checked  the performance of multiplying the ratio of Tink-T predictions (in modified gravity to that in $\Lambda$CDM) to the GM fitting formula for both DGP and $f(R)$. We find that this prediction is comparable to the NBT formula for DGP and does far better overall at modelling the measurements than any other prescription for $f(R)$. \emph{This makes this GM-corrected formula the current state of the art model for  gravity-general, matter bispectrum predictions.} We also note that this prescription is more general than just modifications to gravity, but can encompass non-standard dark energy models too. All that needs to be modified is spherical collapse (\cc{see appendix~\ref{app:sphercol}}).
\\
\\
For the power spectrum, the reaction approach adopted in~\cite{Cataneo:2018cic}  provides a combined perturbation theory-halo model inspired correction to GR-specific non-linear power spectra, which is more sophisticated that the simple ratio of halo model predictions. This results in sub $10\%$ accuracy\footnote{Only $f(R)$ exits the $5\%$ accuracy regime at high redshift and small scales. This can be improved by providing a better non-linear prescription than halofit \cite{Cataneo:2018cic,Giblin:2019iit}.} in the power spectrum for all models considered, at all redshifts and all scales. Such a reaction approach has not been developed for 3-point statistics yet, which becomes motivated by our results, and constitutes a timely endeavour with stage IV surveys already beginning to take data. The reaction method is fast and accurate and provides a promising avenue for improvement in  this sphere of study. We leave this development for future work.
\\
\\
In a secondary analysis, we investigate the impact of non-linearities and inaccuracies in the matter bispectrum on the weak lensing convergence bispectrum for the equilateral configuration. This quantity is closer to what we actually observe from real surveys and so is more relevant to study. For GR and DGP we use the fitting formulae as a benchmark in accuracy. We summarise the results in table~\ref{result_bltab_total}. For both models of gravity, the Tink-T maintains a constant $\sim10\%$ discrepancy with the fitting formulae whereas the Tink-L shows up to $25\%$ discrepancy with the fitting formulae at small multipoles. It does become more accurate than the Tink-T prediction for scales above $\ell \sim 650$. This reflects the large inaccuracies of the Tink-L at low redshift at small $k$, making it largely unsuitable for use in lensing. For $f(R)$ gravity we have no overarching accurate matter bispectrum prescription and so we do no such comparisons. 
\\
\\
We also compare  the ratio of the modified gravity convergence bispectra predictions to that in GR with the GM formula applied. A similar study was conducted using the NBT formula in~\cite{Namikawa:2018erh,Namikawa:2018bju} for the CMB lensing bispectrum with the main conclusion that Horndeski theories with no scale dependent potential term (as assumed for the NBT fitting formula) will give little to no observational signal in upcoming CMB experiments. This assumes that the growth factor is the same as in $\Lambda$CDM, but in practice this is degenerate with $\sigma_8$. Using a combination of clustering and weak lensing measurements, this degeneracy can be partly broken. We find that the signals of modified gravity in the lensing bispectrum vary significantly, relative to \emph{Euclid} like error bars. This of course depends on the modelling choice that is adopted for the matter bispectrum. More accurate models of the matter bispectrum in modified gravity models are therefore needed, which will be the subject of future work. Despite the gravity model parameters we have chosen being ruled out by current observations, there seems to be some prospect for the weak lensing bispectrum to provide a useful probe into modified gravity signatures. 
\\
\\
{\cb On this note, we remark that although the bispectrum generally has a larger sensitivity to modifications than the power spectrum (see Fig.1 of \cite{Bose:2018orj} versus Fig.7 of \cite{Bose:2018zpk} for example), the power of the bispectrum to constrain modifications to $\Lambda$CDM will greatly depend on the survey specifications. It will be detected with a lower significance than the power spectrum, but for GR, it has been shown to add a significant amount of information \cite{Byun:2017fkz,Yankelevich:2018uaz}. Our figure~\ref{bl_all} suggests it can be used to constrain the theories considered here given accurate modelling. Further, the bispectrum earns merit by being able to break degenerecies present in a power spectrum only analysis. For example, \cite{GilMarin:2011xq} suggests that the bispectrum will be a useful tool to break galaxy bias degenerecies in theories beyond $\Lambda$CDM, specifically $f(R)$.}
\\
\\
In future work we aim to provide a more accurate, general and computationally efficient prescription for the \cc{ matter bispectrum at small physical scales}, that improves upon the corrected GM formula \cc{see (eq.~\eqref{eq:correctedgm})} we have tested. We also await ray-tracing simulation measurements in modified gravity to further investigate signals of deviations to GR in weak lensing statistics. Further, various improvements to the halo model ingredients for beyond $\Lambda$CDM  models have  been the  focus of some recent works \cite{Mitchell:2019qke,Arnold:2019zup,Cataneo:2016iav}. Testing these improvements at the bispectrum level will also be left to future work. 
\begin{table}[h]
\centering
\caption{{\bf Summary of matter bispectrum model accuracy:} We give the rough scales in $k[h/{\rm Mpc}]$ at which the models considered are within $10\%$ of the $N$-body measurement over  redshifts in the range  $0 \leq z \leq 1.5$.
Here $k_{\rm l}=0.07h/{\rm Mpc}$ and $k_{\rm h} = 4.05h/{\rm Mpc}$ denote the largest and smallest scales at which we make a measurement from the simulations. \cc{For completeness we also include the tree level and 1-loop results (see appendix~\ref{app:pertvsnl}). See section~\ref{sec:bispectrum} for details on where these scales are deduced, and table~\ref{model_table_1} and table~\ref{model_table_2} for a summary of acronyms}.
}
\begin{tabular}{| c | c | c | c | c | c |}
    \multicolumn{6}{c}{{\bf GR}} \\ \hline
  z  & Tree & 1-loop  & GM/GM-C/NBT  & Tink-T & Tink-L  \\
 \hline
  $0.2$  & $[k_{\rm l},0.08]$  & $[k_{\rm l},0.25]$  & $[k_{\rm l},k_{\rm h}]$  & $[k_{\rm l},2.50]$ & $[k_{\rm l},0.08],[1.00,2.55]$  \\ \hline 
  $0.5$  & $[k_{\rm l},0.15]$  & $[k_{\rm l},0.25]$  & $[k_{\rm l},k_{\rm h}]$  & $[k_{\rm l},0.30],[1.00,3.00]$ & $[k_{\rm l},0.08],[0.70,3.00]$  \\ \hline 
  $1.0$  & $[k_{\rm l},0.15]$  & $[k_{\rm l},0.30]$  & $[k_{\rm l},k_{\rm h}]$  & $[k_{\rm l},0.15],[2.00,3.00]$ & $[k_{\rm l},0.20],[0.50,3.20]$  \\ \hline 
  $1.5 $  & $[k_{\rm l},0.15]$  & $[k_{\rm l},0.38]$  & $[k_{\rm l},1.00]$, $[3.00,k_{\rm h}]$  & $[k_{\rm l},0.22]$ & $[k_{\rm l},3.25]$  \\ \hline 
      \multicolumn{6}{c}{{\bf F5}} \\ \hline
  $0.2$  & $[k_{\rm l},0.15]$  & $[k_{\rm l},0.30]$  & $[k_{\rm l}, 0.08]$ &$[k_{\rm l},2.50]$ & $[k_{\rm l},0.08],[2.00,3.00]$  \\ \hline 
  $0.5$  & $[k_{\rm l},0.15]$  & $[k_{\rm l},0.35]$  & $[k_{\rm l}, 0.15], [0.80,2.50]$  &$[k_{\rm l},3.20]$ & $[k_{\rm l},0.08],[2.00,3.50]$  \\ \hline 
  $1.0$  & $[k_{\rm l},0.15]$  & $[k_{\rm l},0.30]$  & $[k_{\rm l}, 0.40],[0.90,3.20]$& $[k_{\rm l},0.30],[0.90,3.50]$ & $[k_{\rm l},0.20],[1.00,3.80]$  \\ \hline 
  $1.5 $  & $[k_{\rm l},0.15]$  & $[k_{\rm l},0.30]$  &$[k_{\rm l}, k_{\rm h}]$ & $[k_{\rm l},0.15]$, $[1.70,k_{\rm h}]$ & $[k_{\rm l},k_{\rm h}]$  \\ \hline 
\multicolumn{6}{c}{{\bf DGP}} \\ \hline
 $0.0$  & $[k_{\rm l},0.08]$  & $[k_{\rm l},0.08]$  & $[k_{\rm l},0.08],[0.20,1.75]$  & $[k_{\rm l},0.08],[0.20,k_{\rm h}]$ & $[k_{\rm l},0.08],[1.50,k_{\rm h}]$  \\ \hline 
  $1.0$  & $[k_{\rm l},0.15]$  & $[k_{\rm l},0.30]$  & $[k_{\rm l},2.50]$  & $[k_{\rm l},0.15],[1.35,k_{\rm h}]$ & $[k_{\rm l},0.08],[0.55,k_{\rm h}]$  \\ \hline 
  \end{tabular}
\label{result_bstab_total}
\end{table}

\begin{table}[h]
\centering
\caption{{\bf Summary of lensing spectrum accuracy:} We give the rough scales in multipole $\ell$ at which the halo models considered are within $10\%$ of the GM/NBT fitting formula for different source redshifts. Here $\ell_{\rm l}=2$ and $\ell_{\rm h} =2000$ denote the largest and smallest multipoles which we consider.}
\begin{tabular}{| c | c | c | c |}
    \multicolumn{4}{c}{{\bf GR}} \\ \hline
  $z_\star$  & Tree & Tink-T & Tink-L  \\
 \hline
  $1.0$  & $[\ell_{\rm l},100]$  & $[\ell_{\rm l},1400]$  & $[\ell_{\rm l},100],[600,\ell_{\rm h}]$  \\ \hline 
  $1.5$  & $[\ell_{\rm l},100]$  & $[\ell_{\rm l},150],[750,1500]$  & $[\ell_{\rm l},100],[650,\ell_{\rm h}]$  \\ \hline 
    $2.0$  & $[\ell_{\rm l},100]$  & $[\ell_{\rm l},150]$  & $[\ell_{\rm l},100],[750,\ell_{\rm h}]$  \\ \hline 
        \multicolumn{4}{c}{{\bf DGP}} \\ \hline
  $1.0$  & $[\ell_{\rm l},100]$  & $[\ell_{\rm l},1400]$  & $[\ell_{\rm l},100],[600,1900]$  \\ \hline 
  $1.5$  & $[\ell_{\rm l},100]$  & $[\ell_{\rm l},150],[500,1250]$  & $[\ell_{\rm l},100],[700,\ell_{\rm h}]$  \\ \hline 
    $2.0$  & $[\ell_{\rm l},100]$  & $[\ell_{\rm l},200]$  & $[\ell_{\rm l},100],[800,\ell_{\rm h}]$  \\ \hline 
  \end{tabular}
\label{result_bltab_total}
\end{table}

\section*{Acknowledgments}
\noindent  
The authors would like to thank the anonymous referee for useful comments \cb{and improving the clarity of the manuscript}. They also are very grateful to Baojiu Li and Alex Barreira for providing the simulation snapshots from which our measurements were made. The authors also thank Baojiu for useful feedback. The cosmological simulations described in this work were run on the DiRAC Data Centric System at Durham University, United Kingdom, operated by the Institute for Computational Cosmology on behalf of the STFC DiRAC HPC Facility (\url{www.dirac.ac.uk}). This equipment was funded by BIS National E-infrastructure capital grant ST/K00042X/1, STFC capital grants ST/H008519/1 and ST/K00087X/1, STFC DiRAC Operations grant ST/K003267/1 and Durham University. DiRAC is part of the National E-Infrastructure. BB and LL acknowledge support from the Swiss National Science Foundation (SNSF) Professorship grant No.~170547. JB acknowledges support from the SNSF Sinergia grant No.~173716. FL was supported partly by funds of the D\'epartement de Physique Th\'eorique, Universit\'e de Gen\`eve, and partly by a postdoctoral grant from Centre National d'\'Etudes Spatiales (CNES). AMD is supported by the SNSF project, ``The Non-Gaussian Universe and Cosmological Symmetries'', project number: 200020-178787.

\appendix

\section{Perturbation theory in modified gravity}\label{app:modgpt}
In this appendix we describe the perturbative approach to calculate the generalised kernels, $F_i$, used in eq.~\eqref{densitypt}. We solve the continuity and Euler equations order by order
 \begin{eqnarray}
&&a  \delta'(\bfk)+\theta(\bfk) =-
\int\frac{d^3\bfk_1d^3\bfk_2}{(2\pi)^3}\delta_{\rm D}(\bfk-\bfk_{12})
\alpha(\bfk_1,\bfk_2)\,\theta(\bfk_1)\delta(\bfk_2),
\label{eq:Perturb1}\\
&& a \theta'(\bfk)+
\left(2+ \frac{a H'}{H}\right)\theta(\bfk)
-\left(\frac{k}{a\,H}\right)^2\,  \Phi(\bfk)= \nonumber \\ &&
-\frac{1}{2}\int\frac{d^3\bfk_1d^3\bfk_2}{(2\pi)^3}
\delta_{\rm D}(\bfk-\bfk_{12})
\beta(\bfk_1,\bfk_2)\,\theta(\bfk_1)\theta(\bfk_2),
\label{eq:Perturb2}
\end{eqnarray}
where a prime denotes a derivative with respect to the scale factor and $\Phi$ is  the Newtonian   potential. The kernels $\alpha(\bfk_1,\bfk_2)$ and $\beta(\bfk_1,\bfk_2)$ are the standard mode coupling kernels 
\begin{eqnarray}
\alpha(\bfk_1,\bfk_2)=1+\frac{\bfk_1\cdot\bfk_2}{|\bfk_1|^2},
\quad\quad
\beta(\bfk_1,\bfk_2)=
\frac{(\bfk_1\cdot\bfk_2)\left|\bfk_1+\bfk_2\right|^2}{|\bfk_1|^2|\bfk_2|^2}.
\label{alphabeta}
\end{eqnarray}
Modifications to gravity enter through the Poisson equation
\begin{equation}
-\left(\frac{k}{a H(a)}\right)^2\Phi (\bfk;a)=
\frac{3 \Omega_{\rm m}(a)}{2} \mu(k;a) \,\delta(\bfk;a) + S(\bfk;a),
\label{eq:poisson1}
\end{equation}
where $\mu(k;a)$ is the linear modification to GR, while $S(\bfk;a)$ is a source term capturing non-linear modifications, including those responsible for screening effects. The source term is given by 
\begin{eqnarray}
S(\bfk;a)&=&
\int\frac{d^3\bfk_1d^3\bfk_2}{(2\pi)^3}\,
\delta_{\rm D}(\bfk-\bfk_{12}) \gamma_2(\bfk_1, \bfk_2;a)
\delta(\bfk_1)\,\delta(\bfk_2),
\nonumber\\
&& + 
\int\frac{d^3\bfk_1d^3\bfk_2d^3\bfk_3}{(2\pi)^6}
\delta_{\rm D}(\bfk-\bfk_{123})
\gamma_3(\bfk_1, \bfk_2, \bfk_3;a)
\delta(\bfk_1)\,\delta(\bfk_2)\,\delta(\bfk_3) \nonumber \\ 
&& + 
\int\frac{d^3\bfk_1d^3\bfk_2d^3\bfk_3 d^3 \bfk_4}{(2\pi)^9}
\delta_{\rm D}(\bfk-\bfk_{1234})
\gamma_4(\bfk_1, \bfk_2, \bfk_3 , \bfk_4;a)
\delta(\bfk_1)\,\delta(\bfk_2)\,\delta(\bfk_3)\delta(\bfk_4). \nonumber \\
\label{eq:Perturb3}
\end{eqnarray}
The linear $\mu(k;a)$ and higher order $\gamma_i$ modifications to GR can be derived once we specify a particular theory. We refer the reader to \cite{Bose:2016qun,Bose:2018zpk} for the forms of these functions in $f(R)$ and DGP. Further, we note that this framework is very general and can encompass exotic dark energy models too (see \cite{Bose:2017jjx} for example). 
\\
\\
We can now calculate the $F_i$ kernels numerically by solving eqs.~\eqref{eq:Perturb1} and \eqref{eq:Perturb2} order by order, as described in \cite{Taruya:2016jdt,Bose:2016qun,Bose:2018zpk}, and so do not use the analytic forms which can be obtained by using the Einstein-de Sitter approximation as in \cite{Koyama:2009me}.

\section{\cc{Explicit expressions for 1-loop matter statistics}}\label{app:sptexp} 
Here we write out the explicit expressions for the 1-loop bispectrum terms included in eq.~\eqref{1loopps} and eq.~\eqref{1loopbs}. Written explicitly in terms of the integral kernels $F_i$ (see eq.~\eqref{densitypt}) we have \cite{Bose:2018zpk}
\begin{align}
P^{22}(k;z) &= \int \frac{d^3p}{(2\pi)^3} F_2(\bfp,\bfk-\bfp;z)^2 P_0(p)P_0(|\bfk-\bfp|), \\ 
P^{13}(k;z) &= 2F_1(k;z)P_0(k) \int \frac{d^3p}{(2\pi)^3} F_3(\bfp,-\bfp,\bfk;z) P_0(p), 
\end{align}
for the power spectrum, and 
\begin{align}
B^{112}(\bfk_1,\bfk_2,\bfk_3;\,z) & = 2 \Big[ F_2(\bfk_1,\bfk_2;z)F_1(\bfk_1;z)F_1(\bfk_2;z) P_0(k_1) P_0(k_2) + 2 \mbox{perms} (\bfk_1 \leftrightarrow \bfk_2 \leftrightarrow \bfk_3) \Big] \label{tree2} \\ 
B^{222}(\bfk_1,\bfk_2,\bfk_3;\,z) & = 8 \int \frac{d^3p}{(2\pi^3)} F_2(\bfp,\bfk_1-\bfp;z) F_2(-\bfp,\bfk_2+\bfp;z)  F_2(-\bfk_1+\bfp,-\bfk_2-\bfp;z) \nonumber \\ &\times   P_0(p) P_0(|\bfk_1-\bfp|)P_0(|\bfk_2+\bfp|), \label{222comp}\\ 
B^{321-I}(\bfk_1,\bfk_2,\bfk_3;\,z) & = 6 \Big[ F_1(\bfk_1;z)P_0(k_1) \int \frac{d^3p}{(2\pi^3)} F_2(\bfp,\bfk_2-\bfp;z) F_3(-\bfk_1,-\bfp,-\bfk_2+\bfp;z) \nonumber \\ & \times P_0(p) P_0(|\bfk_2-\bfp|)   +  \mbox{5 perms} (\bfk_1 \leftrightarrow \bfk_2 \leftrightarrow \bfk_3) \Big] \\ 
B^{321-II}(\bfk_1,\bfk_2,\bfk_3;\,z) & =6\Big[ F_1(\bfk_1;z) F_2(\bfk_1,\bfk_2;z) P_0(k_1) P_0(k_2) \int \frac{d^3p}{(2\pi^3)}  F_3(\bfk_2,\bfp,-\bfp;z) P_0(p) \nonumber \\ &  +  \mbox{5 perms} (\bfk_1 \leftrightarrow \bfk_2 \leftrightarrow \bfk_3) \Big]\\ 
B^{411}(\bfk_1,\bfk_2,\bfk_3;\,z) & = 12 \Big[F_1(\bfk_1;z) F_1(\bfk_2;z) P_0(k_1) P_0(k_2) \int \frac{d^3p}{(2\pi^3)}  F_4(-\bfk_2, -\bfk_1,\bfp,-\bfp;z) P_0(p) \nonumber \\ &  +  \mbox{2 perms} (\bfk_1 \leftrightarrow \bfk_2 \leftrightarrow \bfk_3)\Big], \label{fourtho}
\end{align}
for the bispectrum, where $\bfk_3 = -\bfk_1 - \bfk_2$. Note that $B^{321} = B^{321-I} + B^{321-II}$ and that here we have just recast the magnitude and angle arguments from eq.~\eqref{1loopbs} in terms of vectors. 

\section{Spherical collapse} \label{app:sphercol}
We follow the Press-Schechter prescription~\cite{Press:1973iz}, which traces the evolution of a spherical top-hat over-density $\delta$, with radius $R_{\rm TH}$ in a homogeneous background spacetime. This evolution is given by mass and momentum conservation equations, yielding~\cite{Schmidt:2008tn}
\begin{equation}
\frac{\ddot{R}_{\rm TH}}{R_{\rm TH}} = -\frac{4\pi G}{3} [\bar{\rho}_m +(1+3w)\bar{\rho}_{\rm eff}] - \frac{1}{3}\nabla^2 \Phi, \label{eq:sphercol}
\end{equation}
where $\bar{\rho}_m$ is the background matter density and $\bar{\rho}_{\rm eff}$ and $w$ are the background energy density and equation of state of an effective dark energy component respectively. In the modified gravity theories considered in this paper $\bar{\rho}_{\rm eff} = \bar{\rho}_\Lambda$ is the energy density of the cosmological constant, and $w=-1$. As in eq.~\eqref{eq:poisson1}, the modifications enter through the Poisson equation 
\begin{equation}
\nabla^2 \Phi = 4\pi G (1+\mathcal{F})\bar{\rho}_{\rm m} \delta,
\label{eq:poisson2}
\end{equation}
with $\mathcal{F}$ depending on the theory of gravity (see Appendix A of \cite{Cataneo:2018cic} for the forms in DGP and $f(R)$ gravity, with $\mathcal{F}=0$ for GR). 
\\ 
\\ 
The over-density  evolves with the top hat as 
\begin{equation}
    \delta = \left( \frac{R_i}{R_{\rm TH}(a)} \right)^3 (1+\delta_i) -1,
\end{equation}
where $R_i$ and $\delta_i$ are the initial top-hat radius and over-density respectively. For a given time, $a_{\rm col}$, we look to find the $\delta_i$ that gives us gravitationally collapsed objects at that time ($R_{\rm TH}(a_{\rm col}) = 0$). We can then approximate the over-density field at the time of collapse by using linear theory $\delta_c(a) = D(a) \delta_i / a_i$ where $a_i$ is the initial scale factor and $D(a)$ is the first order perturbation theory kernel (see eq.~\eqref{densitypt}) in $\Lambda$CDM.
Note that this is an effective quantity for the modified gravity models.
\\
\\
In reality, collapse of over-densities is mixed together with the process of virialisation by which these over-densities become stable bound objects, i.e. halos. One can solve the virial theorem  including any modified gravity or dark energy contributions to obtain the time of virialisation, $a_{\rm vir}$, (see Appendix A of \cite{Cataneo:2018cic}) to get the over-density at the time of virialisation 
\begin{equation}
    \Delta_{\rm vir} = [1+\delta(a_{\rm vir})] \left(\frac{a_{\rm col}}{a_{\rm vir}} \right)^3, 
    \label{deltavir}
\end{equation}
which can then be used to obtain the mass of such a halo assuming sphericity,
\begin{equation}
    M_{\rm vir} = \frac{4\pi}{3} R_{\rm vir}^3 \bar{\rho}_{m,0}\Delta_{\rm vir},
\end{equation}
where $\bar{\rho}_{m,0}$ is the background matter density today. $R_{\rm vir}$ is the corresponding radius of this halo. Using the quantities we have derived here, which are based on some simple assumptions (sphericity of halos, confinement of all matter in halos, etc.), one can begin to construct the matter statistics.

\section{Details of the halo model spectra} \label{app:bias}
\cc{ We here give all details related to the individual terms for the halo model power spectrum and bispectrum  predictions (given in eq.~\eqref{eq:halomps} and eq.~\eqref{eq:halom}). We start by listing the individual terms} 
\begin{align}
P^{\rm 2h}(k) &=   \text{I}_{1}^{1}(k)^2 \ P^{\rm pt}(k), \label{2hps} \\
P^{\rm 1h}(k) &=  \text{I}_{2}^{0}(k,k), \label{eq:12halops} \\
B^{\rm 1h}(k_1,k_2,\mu) &=  \text{I}_{3}^{0}(k_1,k_2,k_3) , \label{eq:1halo} \\
B^{\rm 2h}(k_1,k_2,\mu) &= \text{I}_{2}^{1}(k_1,k_2) \ \text{I}_{1}^{1}(k_3) \ P_L(k_3) + {\rm 2 \; cyclic \; permutations}, \label{eq:2halo} \\
B^{\rm 3h}(k_1,k_2,\mu) &= \text{I}_{1}^{1}(k_1) \ \text{I}_{1}^{1}(k_2) \ \text{I}_{1}^{1}(k_3) \ B^{\rm pt}(k_1,k_2,\mu)  \nonumber \\ 
& + \left[ \text{I}_{1}^{1}(k_1) \ \text{I}_{1}^{1}(k_2) \ \text{I}_{1}^{2}(k_3) \ P_L(k_1) \ P_L(k_2) + {\rm 2 \; cyclic \; permutations} \right] \nonumber \\ 
& + \left[\text{I}_{1}^{1}(k_1) \ \text{I}_{1}^{1}(k_2) \ \text{I}_{1}^{s^2}(k_3) \ S_2(\mathbf{k}_1,\mathbf{k}_2)\ P_L(k_1) \ P_L(k_2) + {\rm 2 \; cyclic \; permutations} \right],
\end{align}
\cc{ where $u(k,M_{\rm vir})$ is the Fourier transform of the halo profile given in eq.~\eqref{eq:nfw}, $k_3 = \sqrt{k_1^2 + k_2^2 + 2k_1 k_2 \mu}$, and we have used the unified notation }
\begin{align}
\text{I}_{\mu}^{\beta}\left(k_1, \dots, k_{\mu} \right) = \int {\rm d} \ln{M_{\rm vir}} \ \left(\frac{M_{\rm vir}}{\bar{\rho}_{m,0}}\right)^{\mu} n_{\rm vir} \left[\prod_{i=1}^{\mu} u\left(k_i,M_{\rm vir}\right)\right] b_{\beta}\left(M_{\rm vir}\right). \label{3hbs}
\end{align}
 The bias terms, $b_{\beta}$, are all discussed at the end of this appendix, but we note here that $b_0(M_{\rm  vir}) = 1$.  
 \\
 \\
 The last term in $B^{\rm 3h}$ is the tidal tensor bias term \cite{McDonald:2009dh}. We use the local Lagrangian assumption \cite{Chan:2012jj,Baldauf:2012hs} to rephrase $\text{I}_{1}^{s^2}(k)$ in terms of $\text{I}_{1}^{1}(k)$,
 \begin{equation}
     \text{I}_{1}^{s^2}(k) = -\frac{4}{7}[\text{I}_{1}^{1}(k) - 1],
 \end{equation}
 and the $S_2$ function is given in GR as 
 \begin{equation}
     S_2(\bfk_1,\bfk_2) = \mu^2 -\frac{1}{3},
\end{equation}
where we remind the reader that $\mu = (\hat{\bfk}_1\cdot \hat{\bfk}_2)$.
\\
\\
 We checked that the term involving $S_2 $ is negligible in GR for the dark matter field, and so we neglect it in our calculations. Note that in the large scale limit this term is the same as in DGP with an overall rescaling of $\mathcal{O}\sim 1$, whereas for $f(R)$ there are additional scale dependencies. We do not expect these to boost this term significantly and so we also neglect it from our modified gravity calculations. Again, all integrations are performed in the range of $ 5 \leq {\rm log_{10}}M_{\rm vir} \leq 18$.
  \\
  \\
Regarding halo bias, if we consider a tree level calculation in the bispectrum, we must consistently consider the bias expansion up to second order. The bias terms are given as (for the Sheth-Tormen and Tinker mass functions respectively) \cite{Sheth:2001dp,Tinker:2010my,Lazanu:2015rta}
\begin{align}
    b_1^{\rm ST}(\nu) &= 1 + \frac{q\nu^2-1}{\delta_c} + \frac{2p}{\delta_c (1+(q\nu^2)^p)}, \\ 
    b_2^{\rm ST}(\nu) & = 2(1 + a_2)(1-b_1^{\rm ST}(\nu)) + \frac{q \nu^2}{\delta_c} \left(\frac{q\nu^2-3}{\delta_c}\right) + \frac{2p}{\delta_c (1+(q\nu^2)^p)} \left( \frac{1+2p}{\delta_c} + 2 \frac{q\nu^2-1}{\delta_c}  \right),\\
     b_1^{\rm T}(\nu) &= \frac{2\phi}{\delta_c[(\beta \nu^2)^{2\phi}+1]} + \frac{\gamma \nu^4 + \delta_c - 2\eta -1}{\delta_c},\\
     b_2^{\rm T}(\nu) & = \frac{2 ( 42 \gamma \nu^4 \phi + 8 \delta_c \phi - 84 \eta \phi + 42 \phi^2 - 21 \phi)}{21 \delta_c^2 [(\beta \nu^2)^{2\phi} +1 ]} +  \frac{21 \gamma^2 \nu^8 + 8 \gamma \delta_c \nu^4 - 84 \gamma \eta \nu^4 - 63 \gamma \nu^4}{21 \delta_c^2} \nonumber \\ &+ \frac{-16 \delta_c \eta - 8 \delta_c + 84 \eta^2 + 42 \eta}{21 \delta_c^2},
\end{align}
where $a_2=-17/21$ and we list other constants below   
\begin{align}
 q &= 0.75, \qquad \qquad \, \, \, p=0.30,  \qquad \qquad  \quad  \, \, \, \, \,   \beta = 0.589 a^{-0.2}, \nonumber \\ 
  \gamma &= 0.864 a^{0.01} ,\quad \quad \eta= -0.243 a^{-0.27}, \quad \quad \phi = -0.729 a^{0.08}.
\label{bolshoiconst}
\end{align}
We re-emphasize that all these constants are fit to Newtonian, $\Lambda$CDM simulations. Finally, we impose the following conditions in order to maintain consistency with matter statistics at large scales 
\begin{align} 
\int_0^\infty d\ln M_{\rm vir} \frac{M_{\rm vir}}{\bar{\rho}} n_{\rm vir}(M_{\rm vir}) b_1(M_{\rm vir}) &= 1, \\ 
\int_0^\infty d\ln M_{\rm vir} \frac{M_{\rm vir}}{\bar{\rho}} n_{\rm vir}(M_{\rm vir}) b_2(M_{\rm vir}) &= 0. 
\end{align}
Since in practice we do not perform the integral over the entire mass range, we apply the method described in the Appendix A of \cite{Schmidt:2015gwz} to ensure that the above consistency relations are upheld when using the mass range adopted in our calculations, $ 5 \leq {\rm log_{10}}M_{\rm vir} \leq 18$. The low mass range cannot generally be measured due to resolution of simulations/instrumentation. 
\\ 
\\
  Finally,  $P^{\rm pt}(k)$  and $B^{\rm pt}(k_1,k_2,\mu)$ are the perturbation theory matter power spectrum and bispectrum predictions. This should be the linear power spectrum ($P_L(k)$) and tree level bispectrum (eq.~\eqref{tree2}) given we have only included bias up to second order. In the main text, although not fully consistent, we also consider $P^{\rm pt}$ and $B^{\rm pt}$ to be the 1-loop spectra. We assume that inaccuracies in the halo mass function for modified gravity theories will dominate over neglecting higher order bias terms. A full treatment of bias up to fourth order in modified gravity is beyond the scope of this paper but for a full treatment of bias at 1-loop order in GR we forward the reader to \cite{Eggemeier:2018qae}. 

\section{\cc{Details of fitting formulae for the bispectrum}}\label{app:gmform}
Here we give the explicit form of the modified second order kernel for the GM and NBT fitting formulae as well as some details on how it is computed. The non-linear prescription enters eq.~\eqref{fittingbh} through the following functions
\begin{align} 
\bar{a}(k,a) &= \frac{1+[\sigma_8(a)]^{a_6} \sqrt{0.7Q(n(k))} (qa_1)^{n(k)+a_2}}{1+(qa_1)^{n(k)+a_2}}, \\
 \bar{b}(k,a) &= \frac{1+0.2a_3 (n(k)+3)(qa_7)^{n(k)+3+a_8} }{1+(qa_7)^{n(k)+3.5+a_8}}, \\
 \bar{c}(k,a) &= \frac{1+[4.5a_4/(1.5+(n(k)+3)^4) ](n(k)+3)(qa_5)^{n(k)+3+a_9} }{1+(qa_5)^{n(k)+3.5+a_9}}, 
\end{align}
with 
\begin{equation}
 Q(x) = (4-2^x)/(1+2^{x+1}) \qquad \mbox{and} \qquad  n(k) = \frac{d\log{P_L^{\rm no-wiggle}(k')}}{d\log{k'}} |_k. \label{eq:nwiggle}
 \end{equation}
The various other quantities are $q=k/k_{NL}$, where $k_{NL}$ is the scale where non-linearities start to become important, determined by solving $k^3_{NL} P_L(k_{NL})/(2\pi^2) = 1$, and $a_{1-9}$ are constants that are determined by fitting to  $N$-body simulations. We use the values found in~\cite{GilMarin:2011ik}, which are determined from GR simulations. Further, $Q(x)$ is known to be a good choice only for $\Lambda$CDM simulations. In the main text we apply a correction to the GM formula which absorbs some of these inaccuracies for beyond $\Lambda$CDM models (see eq.~\eqref{eq:correctedgm}). 
\\
\\
Note that we treat the spurious oscillations in eq.~\eqref{eq:bfit} that arise from oscillations in $n(k)$ (due to baryon acoustic features) by employing the  no-wiggle spectrum as proposed in eq.~(2.47) of~\cite{delaBella:2017qjy}. This was shown to be effective in~\cite{Bose:2018zpk}. {\cb Note that since the DGP linear power spectrum is approximately a rescaling of the $\Lambda$CDM power spectrum (the linear growth is scale independent), then we can just use the $\Lambda$CDM spectrum to compute $n(k)$.}
\\
\\
We also give the explicit values of $\lambda$ and $\kappa$ appearing in eq.~\eqref{fittingbh} for DGP gravity.
\begin{equation}
\kappa(a)=1,  \hspace{.4in} \lambda(a)= \left( 1-\frac{7}{2}\frac{F_{2,DGP}(a)}{F_1(a)^2}\right),
\end{equation} 
where $F_{2,DGP}(a)$ is the second-order growth factor in DGP.
$F_{2,DGP}(a)$ can be determined from solving the following evolution equation~\cite{Koyama:2009me}
 \begin{equation}
 \mathcal{\hat{L}} F_{2,DGP} (a) = -\frac{H_0^2}{24 \Omega_{rc}} \left(\frac{\Omega_{m,0}}{a^3}\right) F_1(a)^2,
 \end{equation}  
where $\Omega_{m,0}$ is the matter density parameter today. The operator, $\mathcal{\hat{L}}$, is given by 
\begin{equation}
\mathcal{\hat{L}} \equiv  a^2 H^2 \frac{d^2}{da^2} + aH^2 \left( 3+ \frac{a H'}{H} \right)\frac{d}{da} - \frac{8 \pi G \rho_m}{2} \left( 1+ \frac{1}{3\beta} \right), \label{lder}
\end{equation}
with
\begin{equation}
    \beta(a) = 1 + \frac{H}{\Omega_{\rm rc}} \left( 1+ \frac{aH'}{3H} \right),
\end{equation}
$\Omega_{\rm rc} = 1/(2H_0r_c)^2$ and $r_c$ being the cross-over scale, taken as a free parameter of DGP theory.

\section{\cc{Power spectrum reaction}}\label{app:reaction}
In this appendix we present the form of the power spectrum reaction, $\mathcal{R}$, as presented in \cite{Cataneo:2018cic}. This takes the following form
\begin{equation}
\mathcal{R}(k;z) = \frac{ [(1-\mathcal{E})\exp{-k/k_\star} + \mathcal{E}] P_L^{\rm real}(k;z) + P_{\rm 1h}^{\rm real}(k;z)}{P_L^{\rm real}(k,z) + P_{\rm 1h}^{\rm pseudo}(k;z)}, \label{eq:reaction}
\end{equation}
where $P_L^{\rm real}(k;z) = F_1(k;z)P_0(k)$ is the linear power spectrum in the desired theory of gravity or dark energy ($F_1$ being the modified linear growth, see eq.~\eqref{densitypt}). $P_{\rm 1h}^{real}(k;z)$ is given in eq.~\eqref{eq:12halops} and $P_{\rm 1h}^{pseudo}(k;z)$ is the same quantity but with spherical collapse being computed without modifications (so in GR)\footnote{Note we use the ST-T prescription (see section~\ref{sec:halobipo}) to compute the halo model terms in the $\mathcal{R}$ computation, as in the original paper.}, but eq.~\eqref{sigmaeq} being computed using the modified linear power spectrum. 
$\mathcal{E}$ is computed as $\mathcal{E} = P_{\rm 1h}^{\rm real}(0.01;z)/P_{\rm 1h}^{\rm pseudo}(0.01;z)$.  Finally, $k_{\rm \star}$, is given by 
\begin{equation}
    k_{\rm \star} = - \bar{k} \left(\ln \left[ 
    \frac{A(\bar{k};z)}{P_L(\bar{k};z)} - \mathcal{E} \right] - \ln\left[1-\mathcal{E}\right]\right)^{-1} \, , \label{kstar0}
\end{equation}
where 
\begin{equation}
    A(k;z) = \frac{[P^{\rm loop, real}(k;z)+ P_{\rm 1h}^{\rm real}(k,z)]
    [P_{\rm L}(k;z)+ P_{\rm 1h}^{\rm pseudo}(k,z)]}
    {P^{\rm loop, pseudo}(k;z)+ P_{\rm 1h}^{\rm pseudo}(k,z)} -  P_{\rm 1h}^{\rm real}(k,z) \, ,\label{kstar}
\end{equation}
with $\bar{k} = 0.06 \, h/{\rm Mpc}$. $P^{\rm loop, real}$ is the 1-loop SPT prediction for the matter power spectrum (eq.~\eqref{1loopps}) in the modified theory and $P^{\rm loop, pseudo}$ is the same quantity but where we set $\gamma_2 = \gamma_3 (= \gamma_4) = 0$ in eq.~\eqref{eq:Perturb3}. We have dropped the superscript in $P_L$ in eq.~\eqref{kstar0} and eq.~\eqref{kstar} as it is assumed that it is always computed within the theory of dark energy or modified gravity under consideration. 

\section{\cc{Comparisons using perturbative treatments}}\label{app:pertvsnl}
\begin{figure}[htbp!]
\centering
  \includegraphics[width=\textwidth,height=7cm]{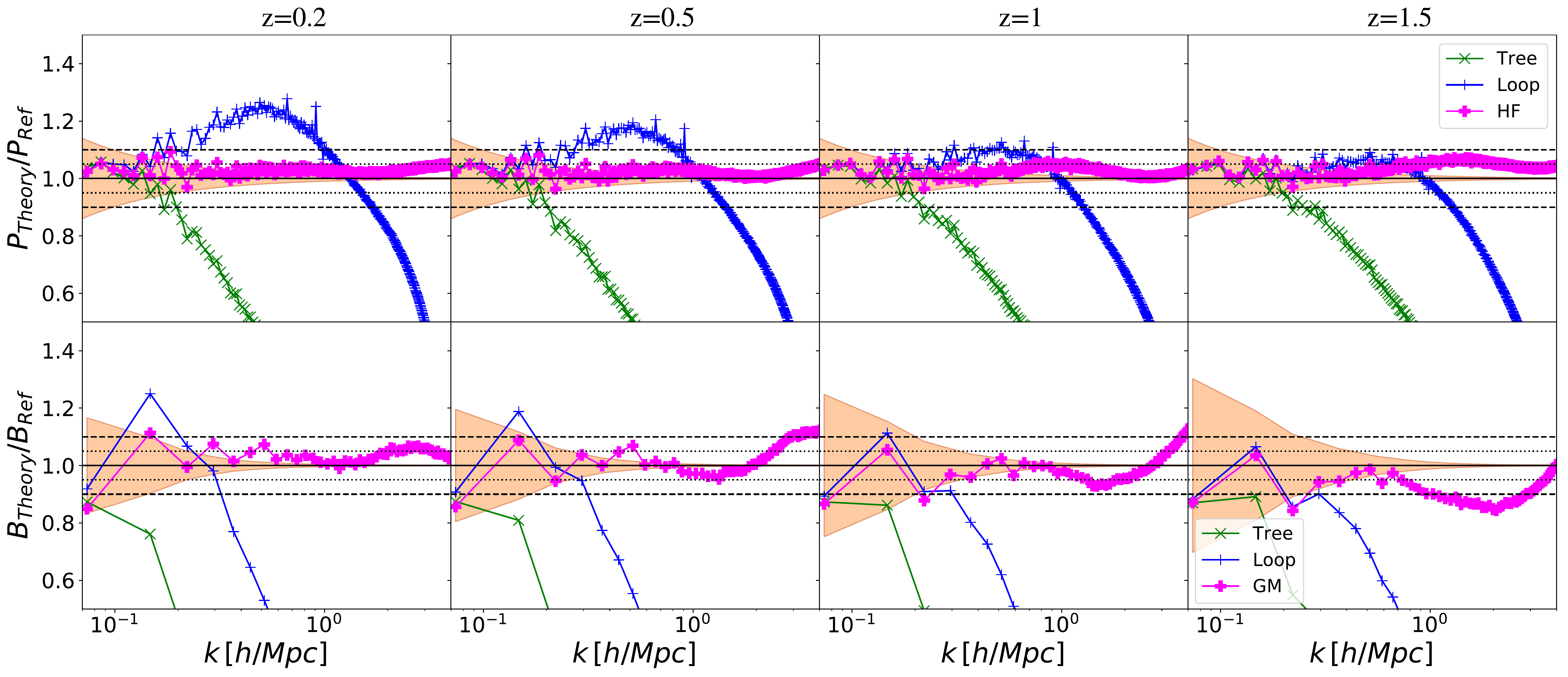}\\
  \caption[CONVERGENCE]{Comparison of the models for the matter power spectrum (top panels) and bispectrum (bottom panels) with $N$-body simulations in GR, from left to right: $z = 0.2, 0.5, 1$.  Tree and  1-loop perturbative predictions are shown as green crosses and blue pluses respectively. The fitting formulae (HF for power spectrum and GM for bispectrum) are shown as magenta full pluses.  The error bands are given by twice eq.~\eqref{pserrs} and eq.~\eqref{bserrs} for power spectrum and bispectrum respectively.}
\label{pert_1}
\end{figure}

\begin{figure}[htbp!]
\centering
  \includegraphics[width=\textwidth,height=7cm]{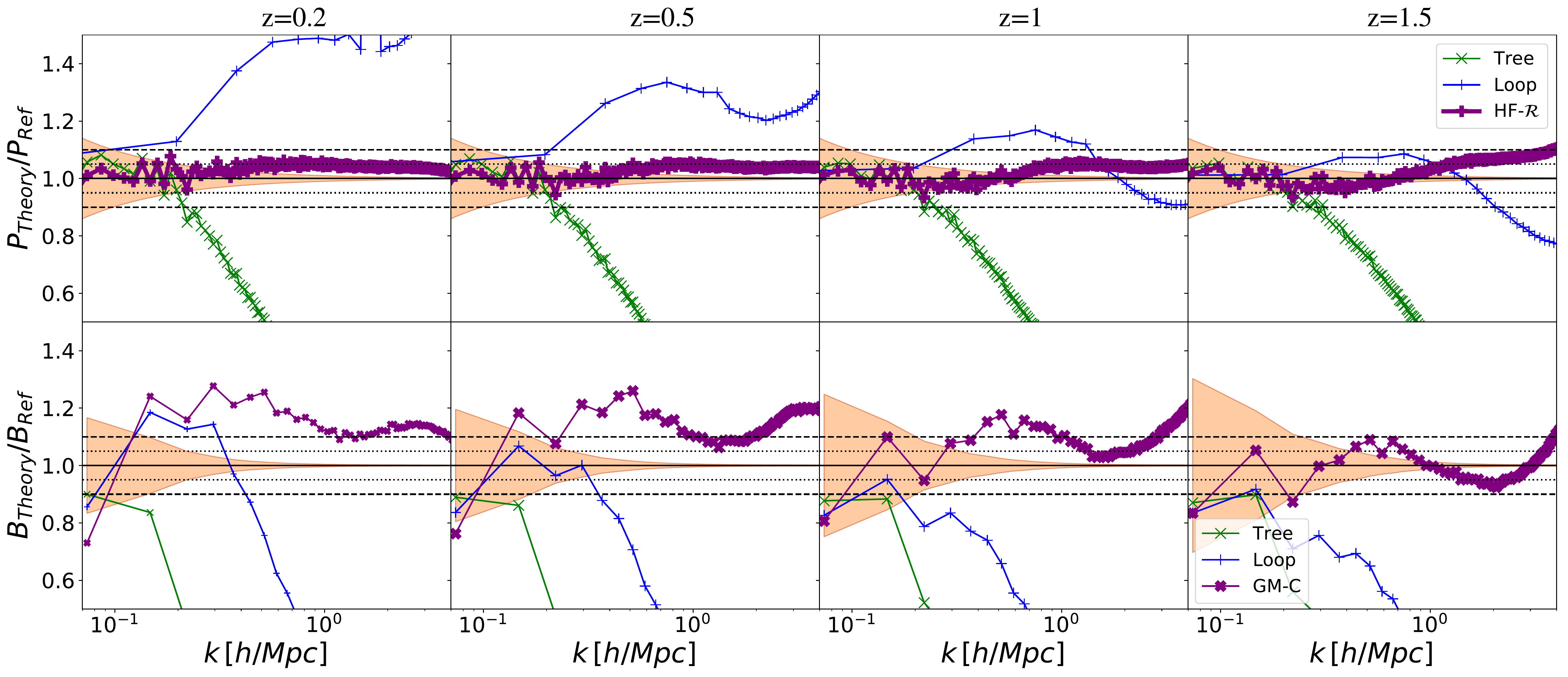}\\
  \caption[CONVERGENCE]{Same as figure~\ref{pert_1} but for $f(R)$ gravity. We include the reaction corrected HF formula and halo-model corrected GM formula, for the power spectrum and bispectrum respectively, as purple full crosses.}
\label{pert_2}
\end{figure}

\begin{figure}[htbp!]
\centering
  \includegraphics[width=\textwidth,height=7cm]{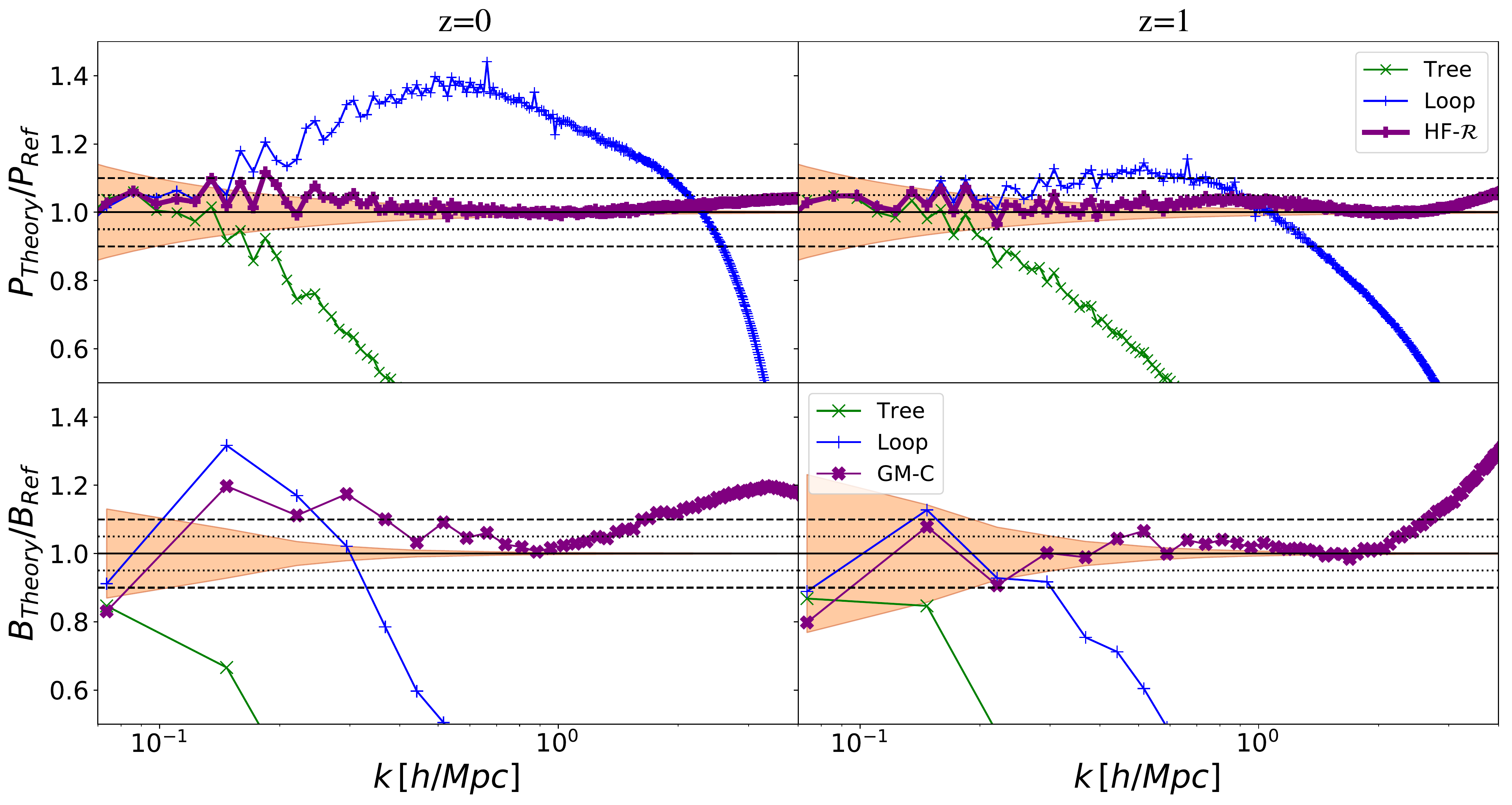}\\
  \caption[CONVERGENCE]{Same as figure~\ref{pert_2} but for DGP gravity.}
\label{pert_3}
\end{figure}

In this appendix we will give some limited comparisons of the tree (eq.~\eqref{tree2}) and 1-loop (eq.~\eqref{1loopps} and eq.~\eqref{1loopbs}) power spectrum and bispectrum predictions against the simulation measurements. We will also include the best non-linear model, as deduced from our main analysis, in these comparisons. For GR this is the halofit and GM formulae for power spectrum and bispectrum respectively. For the modified gravity models, these are the reaction corrected halofit and the GM corrected (eq.~\eqref{eq:correctedgm}) formulae. We show these comparisons in figure~\ref{pert_1}, figure~\ref{pert_2} and figure~\ref{pert_3}.

\bibliography{mybib}
\bibliographystyle{JHEP}
\end{document}